\documentclass[twocolumn,conference,10pt]{IEEEtran}
\makeatletter
\def\ps@headings{%
\def\@oddhead{\mbox{}\scriptsize\rightmark \hfil \thepage}%
\def\@evenhead{\scriptsize\thepage \hfil \leftmark\mbox{}}%
\def\@oddfoot{}%
\def\@evenfoot{}}
\makeatother
\newtheorem{theorem}{Theorem}

\newtheorem{corollary}{Corollary}
\newtheorem{lemma}{Lemma}

\newtheorem{definition}{Definition}

\newcommand{\be}{\begin{equation}}
\newcommand{\ee}{\end{equation}}
\newcommand{\bea}{\begin{eqnarray}}
\newcommand{\eea}{\end{eqnarray}}
\newcommand{\bw}{\begin{eqnarray*}}
\newcommand{\ew}{\end{eqnarray*}}
\newcommand{\bx}{\ensuremath{{\mathbf x}}}

\newcommand{\br}{\ensuremath{{\mathbf r}}}

\newcommand{\D}{\displaystyle}

\usepackage{amssymb,amsmath}
\usepackage{amsmath}
\usepackage{latexsym}
\usepackage{graphicx}
\usepackage{graphics}
\usepackage{subfigure}

\begin{document}

\title{An Axiomatic Theory of Fairness\\ in Network Resource Allocation}

\author{ Tian Lan$^{1}$, David Kao$^{2}$, Mung Chiang$^{1}$, Ashutosh Sabharwal$^{2}$  \\
{  $^{1}$Department of Electrical Engineering, Princeton University, NJ 08544, USA}\\
{  $^{2}$Department of Electrical and Computer Engineering, Rice University, TX 77005, USA}} \maketitle

\maketitle

\begin{abstract}
We present a set of five axioms for fairness measures in resource allocation. A family of fairness measures satisfying the axioms is constructed. Well-known notions such as $\alpha$\mbox{-}fairness, Jain's index, and entropy are shown to be special cases. Properties of fairness measures satisfying the axioms are proven, including Schur-concavity. Among the engineering implications is a generalized Jain's index that tunes the resolution of the fairness measure, a new understanding of $\alpha$-fair utility functions, and an interpretation of ``larger $\alpha$ is more fair''. We also construct an alternative set of four axioms to capture efficiency objectives and feasibility constraints.
\end{abstract}

\section{Quantifying Fairness}

Given a vector $\bx\in{\mathbb{R}}^{n}_{+}$, where $x_{i}$ is the resource allocated to user $i$, \emph{how fair} is it?

One approach to quantify the degree of fairness associated with $\bx$ is through a fairness measure, which is a function $f$ that maps $\bx$ into a real number. Various fairness measures have been proposed throughout the years, e.g., in~\cite{Marson:82, Wong:83, Jain:84, Dianati:05,Koksal:00,Bredel:03}. These range from simple ones, e.g., the ratio between the smallest and the largest entries of $\bx$, to more sophisticated functions, e.g., Jain's index and the entropy function. Some of these fairness measures map $\bx$ to normalized ranges between 0 and 1, where 0 denotes the minimum fairness, 1 denotes the maximum fairness, often corresponding to an $\bx$ where all $x_{i}$ are the same, and a larger value indicate more fairness. For example, min-max ratio~\cite{Marson:82} is given by the maximum ratio of any two user's resource allocation, while Jain's index~\cite{Jain:84} computes a normalized square mean. How are these fairness measure are related? Is one measures ``better'' than any other? What other measures of fairness may be useful?

An alternative approach that has gained attention in the networking research community since~\cite{Kelly:97,Mo:00} is the approach of $\alpha$-fairness and the associated utility maximization.  Given a set of feasible allocations, a maximizer of the $\alpha$-fair utility function satisfies the definition of $\alpha$-fairness. Two well-known examples are as follows: a maximizer of the log utility function ($\alpha=1$) is proportionally fair, and a maximizer of the $\alpha$-fair utility function with $\alpha\rightarrow \infty$ is max-min fair. More recently, $\alpha$-fair utility functions have also been connected to divergence measures~\cite{Masato:09}, and in~\cite{Bonald:01,Massoulie:02}, the parameter $\alpha$ was viewed as a fairness measure in the sense that a fairer allocation is one that is the maximizer of an $\alpha$-fair utility function with larger $\alpha$ --- although the exact role of $\alpha$ in trading-off fairness and throughput can sometimes be surprising~\cite{Tang:06}. While it is often held that $\alpha\rightarrow\infty$ is more fair than $\alpha=1$, which is in turn more fair than $\alpha=0$, it remains unclear what it means to say that $\alpha=3$ is more fair than $\alpha=2$.

Clearly, these two approaches for quantifying fairness are different. On the one hand, $\alpha$-fair utility functions are continuous and strictly increasing in each entry of $\bx$, thus its maximization results in Pareto optimal resource allocations. On the other hand, scale-invariant fairness measures (ones that map $\bx$ to the same value as a normalized $\bx$) are unaffected by the magnitude of $\bx$, and an allocation that does not use all the resources can be as fair as one that does. Can the two approaches be unified?

To address the above questions, we develop an axiomatic approach to measure fairness. We discover that a set of five axioms, each of which simple and intuitive, can lead to a useful family of fairness measures. The axioms are: the Axiom of Continuity, of Homogeneity, of Asymptotic Saturation, of Irrelevance of Partition, and of Monotonicity. Starting with these five axioms, we can \emph{generate} a family of fairness measures from a generator function $g$: any increasing and continuous function that leads to a well-defined ``mean'' function, i.e., any Kolmogorov-Nagumo function \cite{Renyi:60}. For example, using power functions with exponent $\beta$ as the generator function, we derive a unique family of fairness measures $f_{\beta}$ that includes all of the following as special cases, depending on the choice of $\beta$: Jain's index, maximum or minimum ratio, entropy, and $\alpha$-fair utility, and reveals new fairness measures corresponding to other ranges of $\beta$.

In particular, for $\beta\le 1$, well-known fairness measures (e.g., Jain's index and entropy) are special cases of our construction, and we generalize Jain's index to provide a flexible tradeoff between ``resolution'' and ``strictness'' of the fairness measure. For $\beta \ge 0$, $\alpha$-fair utility functions can be factorized as the product of two components: our fairness measure with $\beta=\alpha$ and a function of the total throughput that captures the scale, or efficiency, of $\bx$. Such a factorization also quantifies a tradeoff between fairness and efficiency in achieving Pareto dominance with the maximum possible $\alpha$, and facilitates a clearer understanding of what it means to say that a larger $\alpha$ is ``more fair'' for general $\alpha\in[0,\infty)$.

The axiomatic construction of fairness measures also illuminates their engineering implications. Any fairness measure satisfying the five axioms can be proven to have many useful properties, including Schur-concavity~\cite{Olkin:79}. Consequently, any operation balancing resources between two user always results in a higher fairness value, extending previous results using majorization to characterize fairness~\cite{Dianati:05,Geolt:01}.

The development of an axiomatic theory of fairness takes another turn towards the end of the paper. By removing the Axiom of Homogeneity, we propose an alternative set of four axioms, which allows efficiency of resource allocation be jointly captured in the fairness measure. We show how this alternative system connects with constrained optimization based resource allocation, where magnitude matters due to the feasibility constraint and an objective function that favors efficiency.

The rest of this paper is organized as follows: The five axioms for fairness measures are introduced and discussed in Section~II. Schur-concavity and other properties are proven for any fairness measure satisfying the five axioms in Section~III. We construct a unique family of fairness measures in Section~IV and discuss its relation to previous work. Generalized Jain's index is revealed from this family of fairness measures in Section~V. Section~VI provides a new understanding of $\alpha$\mbox{-}fairness by establishing a connection of our fairness measure to the $\alpha$-fair utility functions. In Section~VII we propose a second set of axioms that directly incorporates a notion of efficiency. Concluding remarks are made in Section~VIII. Due to space limitations all proofs can be found in the online full version~\cite{Tian:09}, together with a discussion contrasting this paper with the well-known axiomatic theories of Nash bargaining solution and Shapley value in economics. Main notation is shown in Table~I. %the Appendices.

%\vspace{0.03in}
\begin{table}[th!]
\begin{center}
\begin{tabular}{cc}
\hline\hline \\[-0.7ex]
{\bf Variable} & {\bf Meaning} \\ \hline \\[-0.7ex]
${\bf x}$ & Resource allocation vector of length $n$ \\[2ex]
${\bf x}^{\uparrow}$ & Sorted vector with smallest element being first \\[2ex]
$w({\bf x})$ & Sum of all elements of ${\bf x}$  \\[2ex]
$f(\cdot),f_{\beta}(\cdot)$ & Fairness measure (of parameter $\beta$)\\[2ex]
$g(\cdot)$ & Generator function \\[2ex]
$s_i$ & Positive weights for weighted mean  \\[2ex]
${\bf 1}_n$ & Vector of all ones of length $n$ \\[2ex]
${\bf x}\succeq {\bf y}$ & Vector ${\bf x}$ majorizes vector ${\bf y}$  \\[2ex]
$\beta$ & Parameter for power function $g(y)=y^\beta$  \\[2ex]
$U_\alpha(\cdot)$ & $\alpha$-fair utility with parameter $\alpha$  \\[2ex]
$H(\cdot)$ & Shannon entropy function  \\[2ex]
$J(\cdot)$ & Jain's index  \\[2ex]
$\Phi_{\lambda}(\cdot)$ & Our utility for fairness and efficiency  \\[2ex]
\hline\hline
\end{tabular} \caption{Table of main notation.}
\end{center}
\vspace{-0.2in}
\end{table}

\section{Axioms}

Let ${\bf x}$ be a resource allocation vector with $n$ non-negative elements. A fairness measure $f({\bf x})$ is a mapping from $\bx$ to a real number, i.e., $f:\mathbb{R}^n_{+}\rightarrow \mathbb{R}$, for all integer $n\ge 1$. We first introduce the following set of axioms about $f$, whose explanations and implications are given next.

\begin{enumerate}

\vspace{0.03in}
\item[{\bf 1)}] \textit{Axiom of Continuity.} Fairness measure $f({\bf x})$ is continuous on $\mathbb{R}^n_{+}$ for all integer $n\ge 1$.

\vspace{0.03in}
\item[{\bf 2)}] \textit{Axiom of Homogeneity.}
Fairness measure $f({\bf x})$ is a homogeneous function of degree 0:
\begin{eqnarray}
f({\bf x})=f(t\cdot {\bf x}), \ \ \forall \ t>0 .
\end{eqnarray}
Without loss of generality, for a single user, we take $|f(x_1)|=1$ for all $x_1>0$, i.e., fairness is a constant for $n=1$.

\vspace{0.03in}
\item[{\bf 3)}] \textit{Axiom of Asymptotic Saturation.}
Fairness measure $f({\bf x})$ of equal resource allocations eventually becomes independent of the number of users:
\begin{eqnarray}
\D \lim_{n\rightarrow\infty} \frac{f{({\bf 1}_{n+1})}}{f({\bf 1}_n)} =1.
\end{eqnarray}

\vspace{0.03in}
\item[{\bf 4)}] \textit{Axiom of Irrelevance of Partition.} If we partition the elements of ${\bf x}$ into two parts ${\bf x}=\left[{\bf x}^{1},{\bf x}^{2}\right]$,  the fairness index $f({\bf x}^{1},{\bf x}^{2})$ can be computed recursively (with respect to a generator function $g(y)$) and is independent of the partition, i.e.,
\begin{equation}
f({\bf x}^{1},{\bf x}^{2})=f\left(w({\bf x}^{1}),w({\bf x}^{2})\right)\cdot  g^{-1}\left( \sum_{i=1}^2 s_i\cdot g\left(f({\bf x}^{i})\right) \right),   \label{a5}
\end{equation}
where $w({\bf x}^{1})$ and $ w({\bf x}^{2})$ denote the sum of resource vectors ${\bf x}^{1}$ and ${\bf x}^{2}$ respectively, and $g(y)$ is a continuous and strictly monotonic function that can generate the following function $h$:
\begin{equation}
h=g^{-1}\left( \sum_{i=1}^2 s_i\cdot g\left(f({\bf x}^{i})\right) \right), \label{g}
\end{equation}
with positive weights satisfying $\sum_i s_i = 1$ such that $h$ qualifies as a \emph{mean} function \cite{Kol:30} of $\{f({\bf x}^{i}), \forall i\}$.

\vspace{0.03in}
\item[{\bf 5)}] \textit{Axiom of Monotonicity.}  For $n=2$ users, fairness measure $f(\theta, 1-\theta)$ is monotonically increasing as the absolute difference between the two elements (i.e. $|1-2\theta|$) shrinks to zero.
\end{enumerate}

Axioms~1--2 are very intuitive. The Axiom of Continuity says that a slight change in resource allocation shows up as a slight change in the fairness measure. The Axiom of Homogeneity says that the fairness measure is independent of the unit of measurement or absolute magnitude of the resource allocation.

Due to the Axiom of of Homogeneity, for an optimization formulation of resource allocation, the fairness measure $f({\bf x})$ alone cannot be used as the objective function if efficiency (which depends on magnitude $\sum_i x_i$) is to be captured. In Section VI, we will connect this fairness measure with an efficiency measure in $\alpha$-fair utility function. In Section VII, we will remove the Axiom of of Homogeneity and propose an alternative set of axioms, which make measure $f({\bf x})$ dependent on both magnitude and distribution of ${\bf x}$, thus capturing fairness and efficiency at the same time.

Axiom 3 is a technical condition used to ensure uniqueness of the fairness measure and invariance under change of variable by fixing a scaling. For example, suppose $f({\bf x})$ is a fairness measure satisfying all axioms (with respect to a mean function $g(y)$) except Axiom 3. It is easy to see that by making a logarithmic change of variables, fairness measure $\log f({\bf x})$ also satisfies all axioms, respect to a mean function $e^{g(y)}$, other than Axiom 3.

\begin{table}[th]
\vspace{0.1in}
\begin{center}
\begin{tabular}{rcc}
\hline
\\[-0.7ex]
Initial: & \multicolumn{2}{c}{$\D \sum_{i=1}^n x_i$} \\[1ex]
 & $\swarrow$ & $\searrow$  \\[1ex]
Level 1: & $\D \sum_{i=1}^k x_i$ & $\D \sum_{i=k+1}^n x_i$  \\[1ex]
 & $\swarrow$ \ \ \ $\searrow$ & $\swarrow$ \ \ \ $\searrow$  \\[1ex]
Level 2: & $x_1,\ldots,x_k$ & $x_{k+1},\ldots,x_n$ \\[0.5ex]
\hline
\end{tabular}\caption{Illustration of the hierarchical computation of fairness.} \label{level}
\end{center}
\vspace{-0.1in}
\end{table}

So far, none of Axioms~1--3 concerns the \emph{construction} of fairness measure as the number of users varies. A hierarchical construction of fairness is used in Axiom 4, which allows us to derive fairness measure $f:\mathbb{R}^n_{+}\rightarrow \mathbb{R}$ of $n$ users recursively from lower dimensions, $f:\mathbb{R}^{k}_{+}\rightarrow \mathbb{R}$ and $f:\mathbb{R}^{n-k}_{+}\rightarrow \mathbb{R}$ for integer $0<k<n$. The recursive computation is illustrated by a two-level representation in Table \ref{level}. Let ${\bf x}^1=[x_1,\ldots,x_k]$ and ${\bf x}^2=[x_{k+1},\ldots,x_n]$. The computation is performed as follows. At level 1, since the total resource is divided into two chunks, $w({\bf x}^1)$ and $w({\bf x}^1)$, fairness across the chunks obtained in this level is measured by $f\left(w({\bf x}^{1}),w({\bf x}^{2})\right)$. At level 2, the two chunks of resources are further allocated to $k$ and $n-k$ users, achieving fairness $f({\bf x}^{1})$ and $f({\bf x}^{2})$, respectively. To compute overall fairness of the resource allocation ${\bf x}=[x_1,x_2,\ldots,x_n]$, we combine the fairness obtained in the two levels using a multiplication in equation (\ref{a5}).

As we consider a continuous and strictly increasing generator function $g(y)$, the function (\ref{g}) is a mean value \cite{Kol:30} for $\{f({\bf x}^{i}),\forall i\}$, which represents the average fairness of individual parts of ${\bf x}$. The set of generator functions giving rise to the same fairness measures may not unique, e.g., logarithm and power functions. The simplest case is when $g$ is identity and $s_{i}=1/n$ for all $i$.
A natural choice of the weight $s_i$ in (\ref{a5}) is to choose the value proportional to the sum resource of vector ${\bf x}^{i}$. More generally, we will consider the following weights
\begin{eqnarray}
s_i=\frac{w^{\rho}({\bf x}^{i})}{\sum_j w^{\rho}({\bf x}^{j})}, \ \forall i \label{w}
\end{eqnarray}
where $\rho\geq 0$ is an arbitrary exponent. When $\rho=0$, weights in (\ref{w}) are equal and lead to an un-weighted mean in Axiom 4. As shown in Section 4, the parameter $\rho$ can be chosen such that the hierarchical computation is independent of partition as stated in Axiom 4. As a special case of Axiom 4, if we denote the resource allocation at level 1 by a vector ${\bf z}=[w({\bf x}^{1}),w({\bf x}^{2})]$ and if the resource allocation at level 2 are equal ${\bf x}^{1}={\bf x}^{2}={\bf y}$, it is straight forward to verify that Axiom 4 implies
\begin{eqnarray}
f({\bf y}\otimes {\bf z})=f({\bf y})\cdot f({\bf z}), \label{direct}
\end{eqnarray}
where $\otimes$ is the direct product of two vectors. As we will show in Section VII, an extension of equation (\ref{direct}) gives an alternative way of stating Axiom 4 and leads to a set of more general axioms on fairness.

Axiom 5 is the only axiom that actually involves a \emph{value} statement on fairness: when there are just two users, more equalized is more fair. This axiom specifies an increasing direction of fairness and ensures uniqueness of $f({\bf x})$. Consider the allocation of a unit resource to two users as ${\bf x}=[\theta,1-\theta]$. It is intuitive that fairness strictly improves as $\theta\rightarrow \frac{1}{2}$, since the difference between the two resource shares tends to be smaller. This intuition also holds for all existing fairness measures, e.g., various, spread, deviation, max-min ratio, Jain's index, $\alpha$-fair utility, and entropy.

By definition, axioms are true, as long as they are consistent and non-redundant. However, not all sets of axioms are useful: unifying known notions, discovering new measures and properties, and providing important insights. We start showing the use of the above five axioms with the following existence (the axioms are consistent) and uniqueness results.
All proofs can be found at~\cite{Tian:09}.
\vspace{0.03in}
\begin{theorem}
(\textit{Existence.}) There exists a fairness measure $f({\bf x})$ satisfying Axioms~1--5.
Furthermore, the fairness achieved by equal-resource allocations ${\bf 1}_n$ is independent of the choice of $g(y)$, i.e.,
\begin{eqnarray}\label{growth}
f({\bf 1}_n)=n^{r}\cdot f(1),
\end{eqnarray}
where $r$ is a constant exponent.
\end{theorem}
\vspace{0.03in}

\begin{theorem}
(\textit{Uniqueness.}) Given a generator function $g$, the resulting $f({\bf x})$ satisfying Axioms~1--5 is unique.
\end{theorem}
%\vspace{0.1in}

\section{Properties of Fairness Measures}

We first prove an intuitive corollary from the five axioms that will be useful for the rest of the presentation.

\vspace{0.03in}
\begin{corollary} \textit{(Symmetry.)}
A fairness measure satisfying Axioms~1--5 is symmetric over ${\bf x}$:
\begin{eqnarray}
f(x_1,x_2,\ldots,x_n)=f(x_{i_1},x_{i_2},\ldots,x_{i_n}),
\end{eqnarray}
where $i_1,\ldots,i_n$ is an arbitrary permutation of indices $1,\ldots,n$.
\end{corollary}
\vspace{0.03in}

The symmetry property shows that the fairness measure $f({\bf x})$ satisfying Axioms~1--5 is irrelevant of labeling of users.

We now make a direct connection of our axiomatic theory to a line of work on measuring statistical dispersion by vector majorization, including the popular Gini Coefficient \cite{Gini:12}. Majorization \cite{Olkin:79} is a partial order over vectors to study whether the elements of vector ${\bf x}$ are less spread out than the elements
of vector ${\bf y}$. We say that ${\bf x}$ is majorized by ${\bf y}$, and we write ${\bf x} \preceq {\bf y}$, if $\sum_{i=1}^n x_i =\sum_{i=1}^n y_i $ (always satisfied due to Axiom 2) and
 \begin{eqnarray}
\sum_{i=1}^d x^{\uparrow}_i \le  \sum_{i=1}^d y^{\uparrow}_i, \ {\rm for} \ d=1, \ldots, n,
 \end{eqnarray}
where $x^{\uparrow}_i$ and $y^{\uparrow}_i$ are the $i$th elements of  ${\bf x}^{\uparrow}$ and ${\bf y}^{\uparrow}$, sorted in ascending order. According to this definition, among the vectors with the same sum of elements, one with the equal elements is the most majorizing vector.

Intuitively, ${\bf x} \preceq {\bf y}$ can be interpreted as ${\bf y}$ being a fairer allocation than ${\bf x}$. It is a classical result \cite{Olkin:79} that ${\bf x}$ is majorized by ${\bf y}$, if and only if, from ${\bf x}$ we can produce ${\bf y}$ by a finite sequence of Robin Hood operations.\footnote{In a Robin Hood operation, we replace two elements $x_i$ and $x_j<x_i$ with $x_i-\epsilon$ and $x_j+\epsilon$, respectively, for some $\epsilon \in (0, x_i-x_j)$. In other words, we take from the rich ($x_i$), and give to the poor ($x_j$).}

Majorization alone cannot be used to define a fairness measure since it is a partial order and fails to compare vectors in certain cases. Still, if resource allocation ${\bf x}$ is majorized by ${\bf y}$, it is desirable to have a fairness measure $f$ such that $f({\bf x})\le f({\bf y})$. A function satisfying this property is known as Schur-concave. In statistics and economics, many measures of statistical dispersion are derived as certain extensions of majorization to the space of non-negative vectors, e.g. Gini Coefficient and Robin Hood Ratio \cite{Gini:12}. We show that our fairness measure is Schur-concave, and therefore can be viewed as a different extension of majorization. Similar to Gini Coefficient and Robin Hood Ratio, the partial order of majorization is preserved by our fairness measure.

\vspace{0.03in}
\begin{theorem}
(\textit{Schur-concavity.}) A fairness measure satisfying Axioms~1--5 is Schur-concave:
\begin{eqnarray}
f({\bf x})\le f({\bf y}), \ {\rm if} \ {\bf x} \preceq {\bf y}.
\end{eqnarray}
\end{theorem}
\vspace{0.03in}

Next we present several properties of fairness measures satisfying the axioms, whose proofs rely on Schur-concavity.

\vspace{0.03in}
\begin{corollary} \textit{(Equal-resource allocation is fairest.)}
A fairness measure $f({\bf x})$ satisfying Axioms~1--5 is maximized by equal-resource allocations, i.e.,
\begin{eqnarray}
\D f({\bf 1}_n)=\max_{{\bf x}\in \mathbb{R}^n} f({\bf x}).
\end{eqnarray}\label{cor:equal}
\end{corollary}
\vspace{0.03in}

\begin{corollary} \textit{(Collecting a fixed-tax is unfair.)}
If a fixed amount $c>0$ of the resource is subtracted from each user (i.e. $x_i-c$ for all $i$), the resulting fairness measure decreases
\begin{eqnarray}
\D f({\bf x} - c\cdot {\bf 1}_n)\le f({\bf x}), \ \ \forall c>0,
\end{eqnarray}
where $c>0$ must be small enough such that all elements of ${\bf x} - c\cdot {\bf 1}_n$ are positive.
\end{corollary}
\vspace{0.03in}

\vspace{0.03in}
\begin{corollary} \textit{(Inactive user achieves no fairness.)}
When a fairness measure $f({\bf x})$ satisfying Axioms~1--5 is generated by by $\rho>0$ in \ref{w},
Removing users with zero resources does not change fairness:
\begin{eqnarray}
\D f({\bf x}, {\bf 0}_n) = f({\bf x}), \ \ \forall n\ge 1.
\end{eqnarray}
\end{corollary}
\vspace{0.03in}

\section{A Family of Fairness Measures}
\subsection{Constructing Fairness Measures}
For any function $g(y)$ satisfying the condition in Axiom 4, we can generate from $g(y)$ a unique $f({\bf x})$. Such an $f({\bf x})$ is a well-defined fairness measure if it also satisfies Axioms~1--5. We then refer to the corresponding $g(y)$ as a generator of the fairness measure.
\vspace{0.03in}
\begin{definition}
Function $g(y)$ is a generator if there exists a $f({\bf x})$ satisfying Axioms~1--5 with respect to $g(y)$.
\end{definition}
\vspace{0.03in}

We note, however, that different generator functions may generate the same fairness measure. Although it is difficult to find the entire set of generators $g(y)$, we have found that many forms of $g(y)$ functions (e.g., logarithm, polynomial, exponential, and their combinations) result in fairness measures equivalent to those generated by the family of power functions. It remains to be determined if all fairness measures satisfying Axioms~1--5 can be generated by power functions.

In this section, we consider power functions, $g(y)=|y|^{\beta}$, parameterized by $\beta$ and derive the resulting family of fairness measures, which indeed satisfy all the axioms. The absolute value ensures that $g(y)$ is non-increasing over $\mathbb{R}_{+}$ for $\beta\ge 0$, and over $\mathbb{R}_{-}$ for $\beta< 0$. From here on, we replace Equation (\ref{a5}) in Axiom 4 by
\begin{eqnarray}
f({\bf x}^{1},{\bf x}^{2})=f\left(w({\bf x}^{1}),w({\bf x}^{2})\right)\cdot  \left( \sum_{i=1}^2 s_i\cdot f^{\beta}({\bf x}^{i}) \right)^{\frac{1}{\beta}}, \nonumber
\end{eqnarray}
where the weights $s_i$ are given by (\ref{w}).

\vspace{0.03in}
\begin{theorem}
(\textit{Fairness measures generated by power functions}) For power mean ($g(y)=|y|^{\beta}$ with parameter $\beta$),  Axioms~1--5 define a unique family of fairness measures as follows
\begin{eqnarray}\label{fairness}
f({\bf x})=\left[\sum_{i=1}^n \left(\frac{x_i}{\sum_j x_j}\right)^{1-\beta r} \right]^{\frac{1}{\beta}}, \ {\rm for}  \ \beta r \le 1
\end{eqnarray}
\begin{eqnarray}\label{fairness2}
f({\bf x})=-\left[\sum_{i=1}^n \left(\frac{x_i}{\sum_j x_j}\right)^{1-\beta r} \right]^{\frac{1}{\beta}}, \ {\rm for}  \ \beta r \ge 1,
\end{eqnarray}
where $r=\frac{1-\rho}{\beta}$ is a constant exponent, which determines the growth rate of maximum fairness as population size $n$ increases, i.e.
\begin{eqnarray}\label{growth}
f({\bf 1}_n)=n^{r}\cdot f(1).
\end{eqnarray}
\end{theorem}
\vspace{0.03in}

For different parameter $\beta$, the fairness measures derived above are equivalent up to a constant exponent $r$: \begin{eqnarray}
f_{\beta,r}({\bf x})=\left[f_{{\beta r},1}\right]^r({\bf x}),
\end{eqnarray}
if we denote $f_{\beta,r}$ as the fairness measure with parameters $\beta$ and $r$. According to Theorem 1, $r$ determines the growth rate of maximum fairness as population size $n$ increases. Without loss of generality, we choose $r=1$ such that the maximum average fairness per user is a constant $\frac{f({\bf 1}_n)}{n}=f(1)$.
From a user's perspective, her perception of maximum fairness is independent of the population size of the system.
From now on, we will use a unified representation of the constructed fairness measurers:
\begin{eqnarray}
f_{\beta}({\bf x})={\rm sign}(1-\beta) \cdot \left[\sum_{i=1}^n \left(\frac{x_i}{\sum_j x_j}\right)^{1-\beta} \right]^{\frac{1}{\beta}}, \label{r0}
\end{eqnarray}
where sign$(\cdot)$ is the sign function.

The special cases are summarized in Table \ref{Table:pre}, where $\beta$ sweeps from $-\infty$ to $\infty$ and $H(\cdot)$ denotes the entropy function. For some values of $\beta$, the corresponding mean function $h$ has a standard name, and for some, known approaches to measure fairness are recovered, while for $\beta\in(0,-1)$ and $\beta\in(-1,-\infty)$, new fairness measures are discovered. For instance, $\beta=-1$ corresponds to a harmonic mean function and results in Jain's index. $\beta=0$ gives a geometric mean function and Shannon entropy. The fairness measure of an arithmetic mean function is discontinuous at $\beta=1$.

\begin{figure}[!th]
 	\centering
 \scalebox{0.45}{\includegraphics[draft=false]{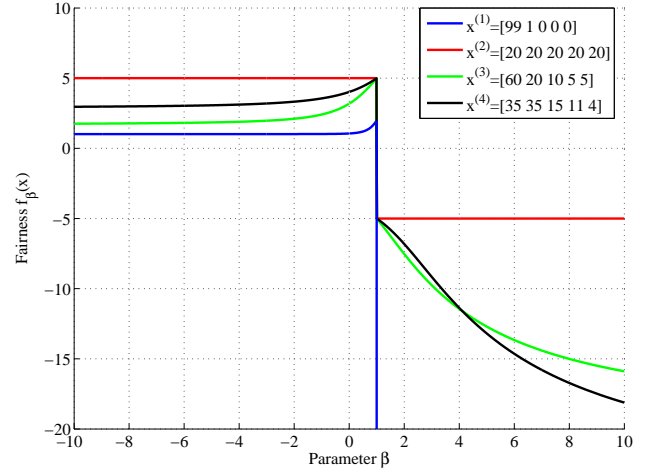}}
	\vspace{-0.05in}
  	\caption{Plot of fairness $f_\beta({\bf x})$ for different values of $\beta$ and sample vectors. The fairness order is is preserved on two sets of sample vectors: $x^{(2)}\succeq x^{(3)} \succeq x^{(1)}$ and $\ x^{(2)} \succeq x^{(4)} \succeq x^{(1)}$. This verifies the Schur-concavity proven in Theorem 1. }
 	\label{fig:sample}
 \end{figure}

To illustrate the fairness measures, we consider an example with $5$ users sharing $100$ units of resource, and plot fairness $f_{\beta}({\bf x})$ over $\beta$ for different sample vectors, i.e., $x^{(1)}=[99 \ 1 \ 0 \ 0 \ 0]$, $x^{(2)}=[20 \ 20 \ 20 \ 20 \ 20]$, $x^{(3)}=[60 \ 20 \ 10 \ 5 \ 5]$, and $x^{(4)}=[35 \ 35 \ 15 \ 11 \ 4]$, in Figure \ref{fig:sample}. The fairness measures are discontinuous at $\beta=0$. For different $\beta$, we also observe that the fairness order is preserved on two sets of sample vectors:
\begin{eqnarray}
& & x^{(2)}\succeq x^{(3)} \succeq x^{(1)} \ {\rm and} \ x^{(2)} \succeq x^{(4)} \succeq x^{(1)} \nonumber
\end{eqnarray}
This is an immediate result from the Schur-concavity proven in Theorem 1. 

%After stating some general results in this section on the above family of fairness measures $f_\beta$, we demonstrate in the following two sections the applications: to generalize Jain's index, and to quantify fairness comparison among alpha-fair utility functions. In Section VII, we will propose a set of more general axioms, from which $\alpha$-fair utility can be directly constructed. Our axiomatic theory provides a unification of many existing fairness indexes and utilities.

%The family of fairness measures in (\ref{r0}) recovers existing forms of both fairness indices (with $\beta<0$) and utility functions (with $\beta>0$). For example, when $\beta=-1$ (i.e., harmonic mean is used in Axiom 4), we get Jain's index $J({\bf x})=f({\bf x})/n$. For $0<\beta<1$ and $\beta>1$, we obtain the part of the $\alpha$-fair utility functions that is related to fairness, as we will show in Section 6 that $\alpha$-fair utility functions are equal to the product of our fairness measure and a function of total throughput for any $\beta=\alpha\ge 0$. Since the $\alpha$-fair utility is discontinuous at $\alpha=1$, the proportional-fair utility can be recovered by making taking the limit of our fairness measure $f_{\beta}$ as $\beta\rightarrow 1$.

\vspace{0.1in}
\begin{table}[th!]
\begin{center}
\begin{tabular}{ccc}
\hline\hline \\[-0.7ex]
 {\bf Value of $\beta$}  & {\bf Our Fairness Measure} & {\bf Known Names} \\ \hline \\[-0.7ex]
{  $\beta\rightarrow\infty$} &  {  $ -\max_i \left\{\frac{\sum_i x_i}{ x_i }\right\}$} & {  Max ratio} \\[2ex]
{  $\beta\in(1,\infty)$} &    {  $-\left[(1-\beta)U_{\alpha=\beta}\left(\frac{{\bf x}}{w({\bf x })}\right) \right]^{\frac{1}{\beta}}$} & {  $\alpha$-fair utility} \\[2ex]
{  $\beta=1$} &  {  $\pm n$ (discontinuous)} & { No name}\\[2ex]
{  $\beta\in(0,1)$} &  {  $\left[(1-\beta)U_{\alpha=\beta}\left(\frac{{\bf x}}{w({\bf x })}\right) \right]^{\frac{1}{\beta}}$} & {  $\alpha$-fair utility}\\[2ex]
{  $\beta\rightarrow 0$}  & {  $e^{H\left(\frac{{\bf x}}{w({\bf x })}\right)} $} & {  Entropy} \\[2ex]
{  $\beta\in(0,-1)$} &  {  $\left[\sum_{i=1}^n \left(\frac{x_i}{w({\bf x})}\right)^{1-\beta r} \right]^{\frac{1}{\beta}}$} & {  No name} \\[2ex]
{  $\beta=-1$} & {  $\frac{(\sum_i x_i)^2}{\sum_i {x_i}^2}=n\cdot J({\bf x})$} & {  Jain's index} \\[2ex]
{  $\beta\in(-1,-\infty)$} &  {  $\left[\sum_{i=1}^n \left(\frac{x_i}{w({\bf x})}\right)^{1-\beta r} \right]^{\frac{1}{\beta}}$} & {  No name} \\[2ex]
{  $\beta\rightarrow -\infty$} &  {  $\min_i \left\{\frac{\sum_i x_i}{ x_i }\right\}$} & {  Min ratio} \\[1ex] \hline\hline
\end{tabular} \caption{Previous results are recovered as special cases of our axiomatic construction. For $\beta\in(0,-1)$ and $\beta\in(-1,-\infty)$, new fairness measures of Generalized Jain's Index are revealed.}\label{Table:pre}
\end{center}
\vspace{-0.1in}
\end{table}

\subsection{Engineering Implications}
%\subsection{Additional Properties}

The fairness measures $f_\beta$ in (\ref{r0}) corresponding to the generator function $g(y)=|y|^{\beta}$ satisfies a number of properties, which give interesting engineering implications to our fairness measure.

%In addition to those proven in Section III, a number of properties of the fairness measures $f_\beta$ in (\ref{r0}) constructed by the generator function $g(y)=y^{\beta}$ are proven.

\vspace{0.03in}
\begin{corollary} \textit{(Number of inactive users.)}
The fairness measures in (\ref{r0}) also count the number of inactive users in the system. When $f_\beta< 0$, $f({\bf x})\rightarrow -\infty$ if any user is assigned zero resource. When $f> 0$,
\begin{eqnarray}
& & \text{Number of users with zero resource}\le n-f({\bf x}), \\
& & \text{Maximum resource to a user}\ge \frac{\sum_i x_i}{f({\bf x})}.
\end{eqnarray}
\end{corollary}
\vspace{0.03in}

\begin{corollary} \textit{(Threshold level of resource.)}
If we increase resource allocation to user $i$ by a small amount $\epsilon$, while not changing other users' allocation, the fairness measures in (\ref{r0}) increases if and only if $x_i < \bar{x}=\left(\frac{\sum_j x_j}{\sum_j x_j^{1-\beta}}\right)^{\frac{1}{\beta}}$ and $0<\epsilon<\bar{x}-x_i$.
\end{corollary}
\vspace{0.03in}

%This corollary implies that $x_f$ serves as a threshold for identifying poor and rich users, since assigning an additional $\epsilon$ amount of resource to user $i$ improves fairness if $x_i<x_f$, while the same assignment reduces fairness if $x_i>x_f$.

\vspace{0.03in}
\begin{corollary} \textit{(Lower bound under box-constraints.)}
If a resource allocation ${\bf x}=[x_1,x_2,\ldots,x_n]$ satisfies box-constraints, i.e., $x_{min}\le x_i\le x_{max}$ for all $i$, the fairness measures in (\ref{r0}) is lower bounded by a constant that only depends on $\beta, x_{min}, x_{max}$:
\begin{eqnarray}
f({\bf x})\ge {\rm sign}(1-\beta) \cdot \frac{\left(\mu\Gamma^{1-\beta}+1-\mu\right)^{\frac{1}{\beta}}}{\left(\mu\Gamma+1-\mu\right)^{\frac{1}{\beta}-1}},
\end{eqnarray}
where $\Gamma=\frac{x_{max}}{x_{min}}$ and $\mu=\frac{\Gamma-\Gamma^{1-\beta}-\beta(\Gamma-1)}{\beta(\Gamma-1)(\Gamma^{1-\beta}-1)}$.
The bound is tight when a $\mu$ fraction of users have $x_{i}=x_{max}$ and the remaining $1-\mu$ fraction of users have $x_{i}=x_{min}$.
\end{corollary}
%\vspace{0.1in}

These results provide intuition on how the family of fairness measures may be interpreted and applied. Through Corollary~5, by specifying a level of fairness, we can limit the number of starved users in a system. Corollary~6 implies that $\bar{x}$ serves as a threshold for identifying ``poor'' and ``rich'' users, since assigning an additional $\epsilon$ amount of resource to user $i$ improves fairness if $x_i<\bar{x}$, and reduces fairness if $x_i>\bar{x}$. Additionally, this provides intuition into threshold methods for allocating resources serially. %Finally, Corollary 7 provides us with a lower bound on evaluation of the fairness measure, with which one might better benchmark empirical values.

\section{Application 1: Generalizing Jain's Index}

When $\beta=-1$ (i.e., harmonic mean is used in Axiom 4), we get a scalar multiple of the widely used Jain's index $J({\bf x})=\frac{1}{n}f({\bf x})$.

Upon inspection of (\ref{r0}) and the specific cases noted in Table~III, we note that any $(-\infty, 0)\cup\beta\in(0,1)$ the range of fairness measure $f_\beta(\bx)$ lies between $1$ and $n$. Equivalently, we can say that the fairness \emph{per user} resides in the interval $\left[\frac{1}{n},1\right]$. When the limit as $\beta\rightarrow 0$ is considered, the resulting fairness measure can also be shown to have this property. Because $f_\beta(\bx)$ for $\beta<1$ has this characteristic, we refer to this subclass of our family of fairness measures as the generalization of Jain's index.

\vspace{0.03in}
\begin{definition}
 $J_\beta({\bf x})=\frac{1}{n}f_\beta({\bf x})$ is a generalized Jain's index parameterized by $\beta\le 1$.
\end{definition}
\vspace{0.03in}

The common properties of our fairness index proven in Section III and IV carry over to this generalized Jain's index. For $\beta=-1$, $J_{-1}({\bf x})$ reduces to the original Jain's index.

%The parameter $\beta$ determines the choice of generator function $g(y)=y^{\beta}$ in Axiom 4. We use $f_{\beta}({\bf x})$ to denote the fairness measures in (\ref{r0}), parameterized by $\beta$. we first prove that, for a given resource allocation ${\bf x}$, fairness $f_{\beta}({\bf x})$ is monotonic as $\beta \rightarrow 1$. Its engineering implication is discussed next.

\begin{figure}[!th]
\begin{center}
\scalebox{0.53}{\includegraphics[draft=false]{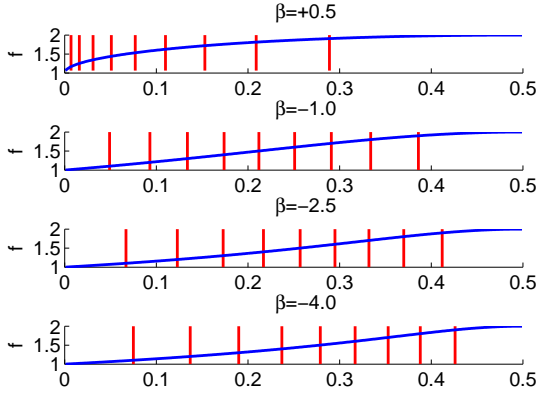}}
\end{center}
\vspace{-0.2in}
\caption{Plot of the fairness measure $f_{\beta}(\theta,1-\theta)$ against $\theta$, for resource allocation ${\bf x}=[\theta,1-\theta]$ and different choices of $\beta=\left\{-4.0,-2.5,-1.0,0.5\right\}$. It can be observed that $f_{\beta}(\theta,1-\theta)$ is monotonic as $\beta \rightarrow 1$. Further, smaller values of $|1-\beta|$ results in a steeper incline over small $\theta$, i.e., the low-fairness region.} \label{fig:beta}
\end{figure}

%\begin{figure*}[!th]
%\begin{minipage}{0.48\textwidth}
%\begin{center}
%\scalebox{0.53}{\includegraphics[draft=false]{alpha.eps}}
%\end{center}
%\end{minipage}
%\hspace{0.1cm}
%\begin{minipage}{0.48\textwidth}
%\begin{center}
%\scalebox{0.49}{\includegraphics[draft=false]{alpha2.eps}}
%\end{center}
%\end{minipage}
%\caption{Plot of the fairness measure $f_{\beta}(\theta,1-\theta)$ against $\theta$, for resource allocation ${\bf x}=[\theta,1-\theta]$ and different choices of $\beta=\left\{-4.0,-2.5,-1.0,0.5\right\}$ and $\beta=\left\{1.5,3.0,4.5,6.0\right\}$, respectively. It can be observed that $f_{\beta}(\theta,1-\theta)$ is monotonic as $\beta \rightarrow 1$. Further, smaller values of $|1-\beta|$ results in a steeper incline over small $\theta$, i.e., the low-fairness region.} \label{fig:beta}
%\end{figure*}

\vspace{0.03in}
\begin{theorem} (\textit{Monotonicity with respect to $\beta$.})
The fairness measures in (\ref{r0}) is negative and decreasing for $\beta\in(1,\infty)$, and positive and increasing for $\beta\in(-\infty,1)$:
\begin{eqnarray}
& & \frac{\partial f_{\beta}({\bf x})}{\partial \beta}\le 0 \ {\rm for} \ \beta\in(1,\infty),  \\
& & \frac{\partial f_{\beta}({\bf x})}{\partial \beta}\ge 0 \ {\rm for} \ \beta\in(-\infty,1).
\end{eqnarray}
%As $\beta \rightarrow 1$, $f$ point-wise converges to constant values:
%\begin{eqnarray}
%\lim_{\beta\uparrow 1} f_{\beta}({\bf x})=n \ {\rm and} \  \lim_{\beta\downarrow 1} f_{\beta}({\bf x})=-n.
%\end{eqnarray}
\end{theorem}
\vspace{0.03in}

%The monotonicity of fairness measures $f_{\beta}({\bf x})$ on $\beta\in (-\infty,1)$ and $\beta\in (1,\infty)$ gives an engineering interpretation of $\beta$.
%Figure \ref{fig:beta} plots fairness $f_{\beta}(\theta,1-\theta)$ for resource allocation ${\bf x}=[\theta,1-\theta]$ and different choices of $\beta=\left\{-4.0,-2.5,-1.0,0.5\right\}$ and $\beta=\left\{1.5,3.0,4.5,6.0\right\}$. The vertical bars in the figure represent the level sets of function $f$, for values $f_{\beta}(\theta_i,1-\theta_i)=\frac{i}{10}\left(f_{max}-f_{min}\right), i=1,2,\ldots,9$. For fixed resource allocations, since $f$ increases as $\beta$ approaches 1, it is observed that the level sets of $f$ are pushed toward the region with small $\theta$ (the region with small fairness values, i.e., the low-fairness region), resulting in a steeper incline in the region. In the extreme case of $\beta=1$, all level set boundaries align with the y-axis in the plot. The fairness measure $f$ point-wise converges to step functions $f_{\beta}(\theta,1-\theta)=2$ and $f_{\beta}(\theta,1-\theta)=-2$, respectively. Therefore, parameter $\beta$ characterized the shape of the fairness measures: a smaller value of $|1-\beta|$ (i.e., $\beta$ closer to 1) causes the level sets to be more condensed in the low-fairness region.

The monotonicity of fairness measures $f_{\beta}({\bf x})$ on $\beta\in (-\infty,1)$ gives an engineering interpretation of $\beta$. Figure \ref{fig:beta} plots fairness $f_{\beta}(\theta,1-\theta)$ for resource allocation ${\bf x}=[\theta,1-\theta]$ and different choices of $\beta=\left\{-4.0,-2.5,-1.0,0.5\right\}$. The vertical bars in the figure represent the level sets of function $f$, for values $f_{\beta}(\theta_i,1-\theta_i)=\frac{i}{10}\left(f_{max}-f_{min}\right), i=1,2,\ldots,9$. For fixed resource allocations, since $f$ increases as $\beta$ approaches 1, the level sets of $f$ are pushed toward the region with small $\theta$ (i.e., the low-fairness region), resulting in a steeper incline in the region. In the extreme case of $\beta=1$, all level set boundaries align with the y-axis in the plot. The fairness measure $f$ point-wise converges to step functions $f_{\beta}(\theta,1-\theta)=2$. Therefore, parameter $\beta$ characterizes the shape of the fairness measures: a smaller value of $|1-\beta|$ (i.e., $\beta$ closer to 1) causes the level sets to be condensed in the low-fairness region.

%While the monotonicty applies for the whole range of $\beta$, it bears a more concrete meaning when the range of the measure is bounded as it is for generalized Jain's index. This corresponds to the left side of Figure~\ref{fig:beta}.
Since the fairness measure must still evaluate to a number between 1 and $n$ here, the monotonicty and resulting shift in granularity of the fairness measure associated with varying $\beta$ suggests differences in evaluating unfairness. At one extreme, $\beta\rightarrow 1$ any solution where no user receives an allocation of zero is fairest. On the other hand, as $\beta\rightarrow -\infty$ the relationship between $f_\beta(\bx)$ and $\theta$ becomes linear, suggesting a stricter concept of fairness --- for the same allocation, as $\beta\rightarrow -\infty$ fairness value drops. Therefore, the parameter $\beta$ can tune the generalization of Jain's index $f$ for different tradeoffs between the resolution and the strictness of fairness measure. %If the fairness measure $f$ is used for classifying different resource allocations, a larger $\beta$ is desirable, since it gives more quantization levels in low-fairness region and provides finer granularity control for unfair resource. On the other hand, if the fairness measure $f$ is used as an objective function, a smaller $\beta$ is desirable, since it has a steeper incline in the low-fairness region and give more incentive for the system to operate in the high-fairness region.

\section{Application 2: Understanding $\alpha$-Fairness}

Due to Axiom 2, the Axiom of Homogeneity, our fairness measures only express desirability over the $(n-1)$-dimension subspace orthogonal to the $\mathbf{1}_n$ vector. Hence, they do not capture any notion of efficiency of an allocation. %The component of resource vectors along the vector $\mathbf{1}_n$ describes another quantity used to classify the efficiency of an allocation, as a function of the sum $w(\bx)$ of resources.

We focus in this section on the widely applied $\alpha$-fair utility function:
\begin{equation}
	\sum_i U_\alpha(x_i), \ {\rm where} \
	U_\alpha(x) = \begin{cases}
			\frac{x^{1-\alpha}}{1-\alpha} & \alpha\geq0,\ \alpha\neq1 \cr
			\log(x) & \alpha = 1\cr
		\end{cases}. \label{eq:aNUM}
\end{equation}
We first show that the $\alpha$-fairness network utility function can be factored into two components: one corresponding to the family of fairness measures we constructed and one corresponding to efficiency. We then demonstrate that, for a fixed $\alpha$, the factorization can be viewed as a single point on the optimal tradeoff curve between fairness and efficiency. Furthermore, this particular point is one where maximum emphasis is placed on fairness while maintaining Pareto optimality of the allocation. This allows us to quantitatively interpret the belief of ``larger $\alpha$ is more fair'' \emph{across all $\alpha\geq 0$}.

\subsection{Factorization of $\alpha$-fair Utility Function}

Re-arranging the terms of the equation in Table~\ref{Table:pre}, we have
\begin{eqnarray}
& U_{\alpha=\beta}({\bf x}) & = \frac{1}{1-\beta}\left|f_{\beta}({\bf x})\right|^{\beta}\left(\sum_i x_i\right)^{1-\beta} \nonumber \\
& & = \left|f_{\beta}({\bf x})\right|^{\beta} \cdot U_{\beta}\left(\sum_i x_i\right), \label{utility}
\end{eqnarray}
where $U_{\beta}\left(\sum_i x_i\right)$ is the one-dimensional version of the $\alpha$\mbox{-}fair utility function with $\alpha=\beta$. For $\beta \rightarrow 1$, it is easy to show that our fairness measure $f_{\beta}(\bx)$, multiplied by a function of throughput $\sum_i x_i$, equals $\alpha$-fair utility function with $\alpha=1$. Similarly, for $\beta \rightarrow\infty$, it equals $\alpha$-fair utility function as $\alpha\rightarrow\infty$. Therefore, Equation (\ref{utility})  also holds for proportional fairness at $\alpha=1$ and max-min fairness at $\alpha\rightarrow \infty$.

Equation (\ref{utility}) demonstrates that the $\alpha$-fair utility functions can be factorized as the product of two components: a fairness measure, $\left|f_{\beta}({\bf x})\right|^{\beta}$, and an efficiency measure, $U_{\beta}\left(\sum_i x_i\right)$. The fairness measure $\left|f_{\beta}({\bf x})\right|^{\beta}$ only depends on the normalized distribution, ${\bf x}/(\sum_i x_i)$, of resources (due to Axiom 2), while the efficiency measure is a function of the sum resource $\sum_i x_i$.

% \vspace{0.03in}
 \begin{table}[th]
 \begin{center}
 \begin{tabular}{rcc}
 \hline
 \\[-0.7ex]
 Allocation: & \multicolumn{2}{c}{${\bf x}$} \\[1ex]
  & $\swarrow$ & $\searrow$  \\[1ex]
 Factorize: & ${\bf x}/\sum_i x_i$ & $\sum_i x_i$  \\[1ex]
  & $\downarrow$ & $\downarrow$  \\[1ex]
 Measure: & $f_{\beta}\left({\bf x}/\sum_i x_i\right)$ & $U_{\beta}\left(\sum_i x_i\right)$ \\[1ex]
 & $\searrow$ & $\swarrow$  \\[1ex]
 Combine: & \multicolumn{2}{c}{$U_{\alpha=\beta}({\bf x})$} \\[0.5ex]
 \hline
 \end{tabular}\caption{Illustration of the factorization of the $\alpha$-fair utility functions into a fairness component of the normalized resource distribution and a efficiency component of the sum resource.} \label{factorization}
 \end{center}
 \vspace{-0.275in}
 \end{table}

The factorization of $\alpha$-fair utility functions is illustrated in Table \ref{factorization} and decouples the two components %from the efficiency component
to tackle issues such as fairness-efficiency tradeoff and feasibility of $\bx$ under a given constraint set.% of utility maximization.
For example, it helps to explain the counter-intuitive throughput behavior in \cite{Tang:06}: an allocation vector that maximizes the $\alpha$-fair utility with a larger $\alpha$ may not be less efficient, because the $\alpha$-fair utility incorporates both fairness and efficiency at the same time.

\subsection{Pareto Optimality in Fairness-Efficiency Tradeoffs}

Although Corollary~\ref{cor:equal} states equal allocation is fairest, an $\alpha$-fair allocation may not have an equal distribution. This is because the additional efficiency component in (\ref{utility}) can skew the optimizer (i.e., the resource allocation resulting from $\alpha$\mbox{-}fair utility maximization) away from an equal distribution. For this to happen there must exist an allocation that is feasible (within the constraint set of realizable allocations) with a large enough gain in efficiency over all equal distribution allocations. Hence, the magnitude of this skewing depends on the fairness parameter ($\alpha=\beta$), the constraint set of $\bx$, and the relative importance of fairness and efficiency.

Guided by the product form of (\ref{utility}), we consider a scalarization of the maximization of the two objectives: fairness and efficiency:
\begin{equation}
	\Phi_{\lambda}(\mathbf{x}) = \lambda \ell\left(f_\beta\left(\mathbf{x}\right)\right) + \ell\left(\sum_i x_i\right),
	\label{eq:tradeoff}
\end{equation}
where $\beta\in(0,1)\cup(1,\infty)$ is fixed, $\lambda\in[0,\infty)$ absorbs the exponent $\beta$ in the fairness component of (\ref{utility}) and is a weight specifying the relative emphasis placed on the fairness, and
\begin{equation}
	\ell(y) = {\rm sign}(y)\log(|y|).
\end{equation}
The use of the log function later recovers the product in the factorization of (\ref{utility}) from the sum in the scalarized (\ref{eq:tradeoff}).

An allocation vector $\mathbf{x}$ is said to be Pareto dominated by $\mathbf{y}$ if $x_i\leq y_i$ for all $i$ and $x_i< y_i$ for at least some $i$. An allocation is called Pareto optimal if it is not Pareto dominated by any other feasible allocation. If the relative emphasis on efficiency is sufficiently high, Pareto optimality of the solution can be maintained.
To preserve Pareto optimality, we require that if $\mathbf{y}$ Pareto dominates $\mathbf{x}$, then $\Phi_{\lambda}(\mathbf{y}) > \Phi_{\lambda}(\mathbf{x})$.

\vspace{0.03in}
\begin{theorem}\textit{Preserving Pareto optimality.}
The necessary and sufficient condition on $\lambda$ such that $\Phi_{\lambda}(\mathbf{y}) > \Phi_{\lambda}(\mathbf{x})$  if $\mathbf{y}$ Pareto dominates $\mathbf{x}$ is
	\begin{equation}
		\lambda \leq \left|\frac{\beta}{1-\beta}\right|.\label{eq:paretocond}
	\end{equation}
	\label{thm:pareto}
\end{theorem}
\vspace{0.03in}

Consider the set of maximizers of  (\ref{eq:tradeoff}) for $\lambda$ in the range in Theorem 6:
\begin{equation}
\mathbb{P}=\bigg\{{\bf x}: {\bf x}={\rm arg}\max_{{\bf x}\in \mathbb{R}} \Phi_{\lambda}(\mathbf{x}), \ \forall \lambda \leq \left|\frac{\beta}{1-\beta}\right|  \bigg\}. \label{optimizer}
\end{equation}
When weight $\lambda=0$, the corresponding points in $\mathbb{P}$ is most efficient. When weight $\lambda=\left|\frac{\beta}{1-\beta}\right|$, it can be shown that the factorization in (\ref{utility}) is equivalent to (\ref{eq:tradeoff}). Therefore, $\alpha$-fairness corresponds to the solution of an optimization that places the maximum emphasis on the fairness measure parameterized by $\beta=\alpha$ while preserving Pareto optimality. Allocations in $\mathbb{P}$ corresponding to other values of $\lambda$ achieve a tradeoff between fairness and efficiency, while Pareto optimality is preserved.

\begin{figure}[!th]
\begin{center}
	\subfigure[]{
\scalebox{0.45}{\includegraphics[draft=false]{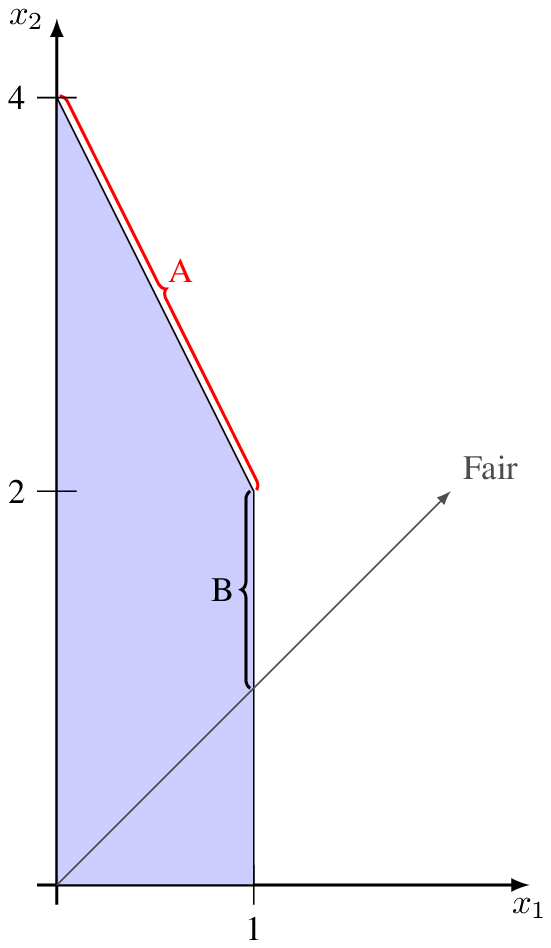}}
	\label{fig:example_a}}
	\subfigure[]{
\scalebox{0.33}{\includegraphics[draft=false]{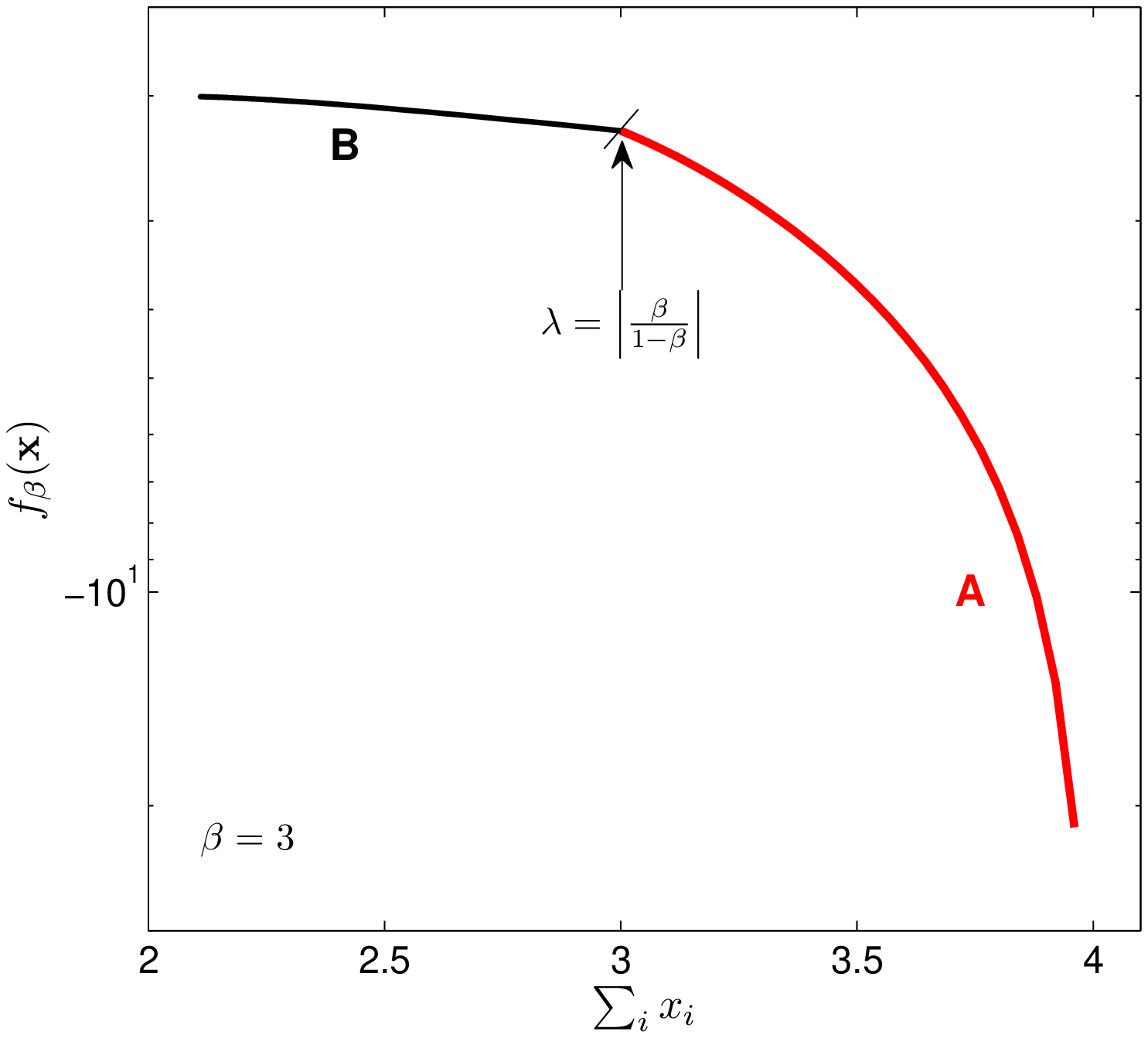}}
	\label{fig:example_b}}
	\caption{\subref{fig:example_a} Feasible region (i.e., the constraint set of the utility maximization problem) where overemphasis of fairness violates Pareto dominance, and \subref{fig:example_b} its fairness-efficiency tradeoff for $\beta=3$. Region A corresponds to Pareto optimal solutions. Region B is when the condition of Theorem 6 is violated, and solutions are more fair, but no longer Pareto optimal.}
	\label{fig:example}
\end{center}
\end{figure}

Figure~\ref{fig:example_b} illustrates an optimal fairness-efficiency tradeoff curve $\bigg\{\big[f_{\beta}({\bf x}),\Sigma_i x_i \big], \ \forall {\bf x}={\rm arg}\max_{{\bf x}\in \mathbb{R}} \Phi_{\lambda}(\mathbf{x}), \ \forall \lambda \bigg\}$ corresponding to the constraint set  shown in Figure~\ref{fig:example_a}. The set of optimizers $\mathbb{P}$ in (\ref{optimizer}), which is obtained by maximizing Pareto optimal utilities (\ref{eq:tradeoff}), is shown by curve $A$ in Figure~\ref{fig:example_b}.

\subsection{Why Larger $\alpha$ is More Fair}

In the previous subsection we demonstrated the factorization (\ref{utility}) is an extreme point on the tradeoff curve between fairness and efficiency for fixed $\beta=\alpha$. What happens when $\alpha$ becomes bigger?

We denote by $\bigtriangledown_\mathbf{x}$ the gradient operator with respect to the vector $\mathbf{x}$. For a differentiable function, we use the standard inner product ($\left\langle\mathbf{x},\mathbf{y}\right\rangle = \sum_i x_iy_i$) between the gradient of the function and a normalized vector to denote the directional derivative of the function.

\begin{theorem}\textit{(Monotonicity of fairness-efficiency reward ratio.)}
	Let allocation $\mathbf{x}$ be given. Define $\boldsymbol{\eta} = \frac{1}{n}\mathbf{1}_n - \frac{\mathbf{x}}{\sum x_i}$ as the vector pointing from the allocation to the nearest fairness maximizing solution. Then the fairness-efficiency reward ratio:
	\begin{equation}
		\frac{\left\langle\displaystyle\bigtriangledown_\mathbf{x} U_{\alpha=\beta}(\mathbf{x}),\frac{\boldsymbol{\eta}}{\|\boldsymbol{\eta}\|}\right\rangle}
		{\left\langle\displaystyle\bigtriangledown_\mathbf{x} U_{\alpha=\beta}(\mathbf{x}),\frac{\mathbf{1}_n}{\|\mathbf{1}_n\|}\right\rangle},
		\label{ratioequation}
	\end{equation}
	is non-decreasing with $\alpha$, i.e. higher $\alpha$ gives a greater relative reward for fairer solutions.
	\label{thm:fairer}
\end{theorem}

The the choice of direction $\boldsymbol{\eta}$ is a direct result of Axiom~2 and Corollary~2, which together imply that $\boldsymbol{\eta}$ is the direction that most increases fairness and is orthogonal to increases in efficiency.

An increase in either fairness or efficiency is a ``desirable'' outcome. The choice of $\alpha$ dictates exactly how desirable one objective is relative to the other (for a fixed allocation).
Theorem~\ref{thm:fairer} states that, with a larger $\alpha$, there is a larger component of the utility function gradient in the direction of fairer solutions, relative to the component in the direction of more efficiency. %Economically, this could be seen as decreasing the pricing of fairer allocations relative to a fixed price on allocations of increased throughput.
Notice, however, that comparison must be in terms of the ratio between these two gradient components rather than the magnitude of the gradient, and both fairness and efficiency may increase simultaneously.

\begin{figure}[!th]
	\begin{center}
  %\centerline{\epsfxsize 1.00\linewidth \epsfbox{monotonic.eps}}
\scalebox{0.48}{\includegraphics[draft=false]{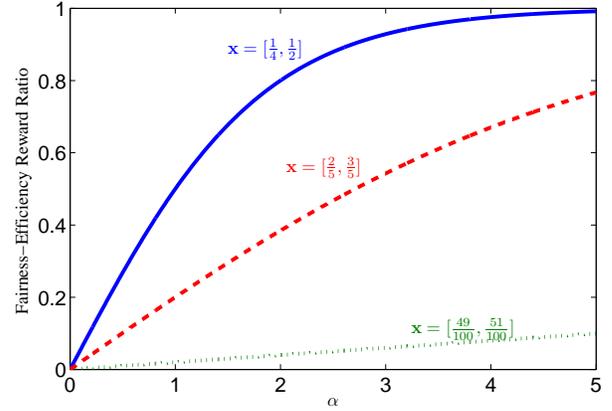}}
	\vspace{-0.3in}\
 	\caption{Monotonic behavior of the ratio (\ref{ratioequation}) as a function of $\alpha$. Three fixed allocations are considered, and solutions that are already more fair have a lower ratio.}
	\label{fig:ratio}
	\end{center}
	\vspace{-0.1in}
\end{figure}

This result provides a justification for the belief that larger $\alpha$ is ``more fair'', not just for $\alpha\in\{0,1,\infty\}$, but for any $\alpha\in[0,\infty)$. Figure~\ref{fig:ratio} depicts how this ratio increases with $\alpha=\beta$ for some examples allocations.

\section{Alternative Axioms}
Given a set of useful axioms, it is important to ask if other useful axiomatic systems are possible. By removing or modifying some of the five axioms here, for example, Axiom~2 that decouples the concern on efficiency from fairness, what kind of fairness measures will result? Can an alternative set of axioms lead to the construction of fairness measures that do not automatically decouple from the notions of efficiency and feasibility of resource allocation?

In this section, we propose a set of alternative axioms, which includes Axioms~1--5 as a special case. Let $F:\mathbb{R}^n_{+}\rightarrow \mathbb{R}$ be a general fairness measure satisfying four axioms as follows.

\begin{enumerate}

\vspace{0.03in}
\item[{\bf 1$^\prime$)}] \textit{Axiom of Continuity.} Fairness measure $F({\bf x})$ is continuous on $\mathbb{R}^n_{+}$ for all integer $n\ge 1$.

\vspace{0.03in}
\item[{\bf 2$^\prime$)}] \textit{Axiom of Asymptotic Saturation.}
Fairness measure $f({\bf x})$ of equal resource allocations eventually becomes independent of the number of users:
\begin{eqnarray}
\D \lim_{n\rightarrow\infty} \frac{F{({\bf 1}_{n+1})}}{F({\bf 1}_n)} =1.
\end{eqnarray}

\vspace{0.03in}
\item[{\bf 3$^\prime$)}] \textit{Axiom of Irrelevance of Splitting.} For an allocation vector ${\bf x}=[x_1,x_2]$, we split each element $x_i$ into multiple elements by a direct product $x_i{\bf y^i}$, where ${\bf y^i}$ is a non-negative vector. If the splitting vectors have equal weights $w({\bf y^1})=w({\bf y^2})$, the fairness of the new allocation vector $[x_1{\bf y^1},x_2{\bf y^2}]$ is given by
\begin{equation}
F(x_1{\bf y^1},x_2{\bf y^2})=F\left({\bf x}\right)\cdot  g^{-1}\left( \sum_{i=1}^2 s_i\cdot g\left(F({\bf y}^{i})\right) \right),   \label{a5b}
\end{equation}
where $\sum_i s_i = 1$ are positive weights and $g(y)$ is a continuous and strictly increasing function.

\vspace{0.03in}
\item[{\bf 4$^\prime$)}] \textit{Axiom of Monotonicity.}  For $n=2$ users, fairness measure $F(x_1, x_2)$ increases as ratio $x_1/x_2$ goes to 1, when sum resource $x_1+x_2$ is fixed.
\end{enumerate}

Axioms~1$^\prime$ and 2$^\prime$ remain the same as Axioms~1 and 3 before. Axiom~4$^\prime$ is equivalent to Axiom~5 with the additional qualification that the sum-resource does not change. This qualification was previously unnecessary due to Axiom~2 --- $f(\bx)$ does not vary with the amount of total resources --- however, is now required in the new set of axioms. Axiom~3$^\prime$ is used to recursively construct fairness measure $F({\bf x})$ from lower dimensions and is similar to the Axiom~4. The vector $[x_1{\bf y^1}, x_2{\bf y^2}]$ can be viewed as a generalized direct product of vector ${\bf x}$ with two different vectors ${\bf y^1}$ and ${\bf y^2}$, which split the resource of each element of ${\bf x }$ to multiple users. If ${\bf y^1}={\bf y^2}$, this splitting reduces to a direct product.

Since the Axiom of Homogeneity is removed, fairness measure $F({\bf x})$ depends on the absolute magnitude of resource vector ${\bf x}$. Using Axiom~3$^\prime$, we can prove that $F({\bf x})$ is a homogeneous function of real degree. Furthermore, the two sets of axioms are equivalent, if the order of homogeneity is zero. This means that the new axiomatic system is more general than the original one.

\vspace{0.03in}
\begin{theorem}
(\textit{Existence and Uniqueness.}) For each generator $g(y)$, there exists a unique fairness measure $F({\bf x})$ satisfying Axioms~1$^\prime$--~4$^\prime$. We have,
\begin{eqnarray}\label{relationship}
F({\bf x})=f({\bf x})\cdot \left(\sum_i x_i\right)^{\frac{1}{\lambda}}
\end{eqnarray}
where $\frac{1}{\lambda}\in \mathbb{R}$ is the degree of homogeneity and $f({\bf x})$ is a fairness measure satisfying Axioms~1--5 with respect to the same generator $g(y)$.
\end{theorem}
\vspace{0.03in}

While it is easy to verify that some properties, like that of symmetry, in Section III also hold for fairness measure $F({\bf x})$, some properties of fairness measures satisfying Axioms~1--5 are lost in the generalization. For instance, we can no longer say that equal allocations are best.

When power generators $g(y)=|y|^\beta$ are considered, from Axioms~1$^\prime$--~4$^\prime$ we can derive fairness measure $F_{\beta,\lambda}({\bf x})$, which is parameterized by both $\lambda$ and $\beta$,
\begin{eqnarray}\label{relationship1}
F_{\beta,\lambda}({\bf x})=f_{\beta}({\bf x})\cdot \left(\sum_i x_i\right)^{\frac{1}{\lambda}}.
\end{eqnarray}
This unifies our results in Sections IV-VI: Generalized Jain's index is a special case of $F_{\beta,\lambda}({\bf x})$ for $1/\lambda=0$ and $\beta<1$; fairness measure $f_\beta({\bf x})$ is a subclass of $F_{\beta,\lambda}({\bf x})$ for $\lambda=0$; and $\alpha$-utility is obtained for $1/\lambda=\beta/(1-\beta)$ and $\beta>0$ by comparing (\ref{relationship1}) and (\ref{eq:tradeoff}). The degree of homogeneity $1/\lambda$ determines how $F_{\beta,\lambda}({\bf x})$ scales as throughput increases. The decomposition of fairness and efficiency in Section VI is now an immediate consequence from Axioms~ 1$^\prime$--~4$^\prime$.

There is a useful connection with the characterization of $\alpha$-fair utility function in the last section. The absolute value $|\lambda|$ is equivalent to the parameter used for defining the utility function (\ref{eq:tradeoff}) in Section VI.B. From Theorem 6, we can conclude that fairness measure $F_{\beta,\lambda}({\bf x})$ is Pareto optimal if and only if
\begin{eqnarray}\label{eq:cond}
\frac{1}{|\lambda|} \ge \left|\frac{\beta}{1-\beta}\right|.
\end{eqnarray}
For every $\beta$, there is a minimum degree of homogeneity such that Pareto optimality can be achieved. When inequality (\ref{eq:cond}) is not satisfied, $F_{\beta,\lambda}({\bf x})$ loses Pareto optimality and produces less throughput-efficient solutions if it is used as an objective function in utility optimization. Fairness measures with small degree of homogeneity $1/\lambda$ are more suitable for computing index values of fairness.

The degree of homogeneity of a fairness measure satisfying Axioms~1$^\prime$--~4$^\prime$ parameterizes a tradeoff between the concept of fairness and efficiency. Moreover, when power functions are used as generating functions, the degree of homogeneity is equivalent to $\frac{1}{\lambda}$ in (\ref{eq:tradeoff}). Therefore, the intuition behind our result on a maximum $|\lambda|$ (minimum degree of homogeneity) to ensure Pareto optimality can be extended to the general optimization-theoretic approach to fairness, i.e. for a fairness measure $F$ generated from any $g$, there is a minimum degree of homogeneity $\frac{1}{\lambda}$ to produce a Pareto optimal solution. Just like our first set of axioms generalized Jain's index and revealed new fairness measures with desirable properties, the second set of axioms offers a rich family of objective functions.

To summarize, in removing Axiom~2 and adapting the set of axioms accordingly, we have shown that Axioms~1$^\prime$--~4$^\prime$ include Axioms~1--5 as special cases. The resulting measures are now affected by notions of both efficiency and fairness, with the balance between the two governed by the degree of homogeneity $\frac{1}{\lambda}$.

\section{Related Axiomatic Theories}
Axiomatic theories often form a stable foundation for wide applications. Famous examples include that of the first order logic and that of Nash equilibrium. In network economics, two prominent axiomatic theories have been used to study network resource allocation. The Nash bargaining solution~\cite{Nash:1950b} --- of which proportional fairness is a generalization and which models the bartering of persons with an initial allotment of goods --- is based on a system of four axioms. In cooperative game theory in the Shapley value solution concept~\cite{Shapley:1953}. Given the setup of a coalitional game, four axioms\footnote{Although usually the axioms are presented as a triple, the Shapley value is the only efficient solution, and thus efficiency of a coalitional strategy can be considered an axiom} uniquely define the Shapley value as the solution concept.

In both of these constructions, effiency was a fundamental axiom in defining the solution. In fact, both these approaches to cooperative allocation hold Pareto optimality as an axiom and thus are more akin to the optimization theoretic approach to fairness.

The first family of fairness measures, $f$, is confined to homogeneous functions of degree zero, and the resulting measures are an integral part of the $\alpha$-fairness utility function. One might suspect that it is possible to extend our axiomatic structure to the optimization theoretic fairness approach by relaxing the Axiom of Homogeneity to homogeneous functions of arbitrary degree. This is indeed the case as developed by the previous section.

\section{Concluding Remarks}

An axiomatic approach to the fundamental concept of fairness illuminates many issues in network resource allocation research. This paper is far from the end of axiomatic \emph{theories} of fairness. One way to re-examine axioms is to refute their corollaries in the context of network resource allocation.  Perhaps all $x_{i}$ being the same should not be a maximizer of fairness measure, and a fairness measure need not be Schur-concave. Instead, making some $x_{i}$ bigger should be called more fair if the resulting $\bx$ is bigger in all coordinates, i.e., those contributing to the overall efficiency should ``fairly'' receive more resources. Perhaps the fairness measure should be a function dependent on the feasible region of allocations. These possibilities mean that alternative sets of axioms of fairness, ones with a value statement different from that in Axiom~5 or Axiom~4$^\prime$ in this paper, deserve further exploration.

We have also assumed that resource is infinitesimally divisible and has no 'user or time dependency', that the way resource allocation is decided (e.g., by a central controller or autonomously) is irrelevant, and that the actual allocation can be transparently verified. None of these assumptions is true. Removing them will further enrich axiomatic theories of fairness in resource allocation.

\appendix

\subsection{Proof of Theorems 1 and 2}

We first show that fairness achieved by equal-resource allocations ${\bf 1}_n$ is independent of the choice of $g(y)$. Without loss of generality, we assume that $f(1)=1$.

\begin{lemma} To satisfy Axioms~1--5, fairness achieved by equal-resource allocations ${\bf 1}_n$ is given by
\begin{eqnarray}\label{growth1}
f({\bf 1}_n)=n^{r}\cdot f(1), \ \ \forall \ n\ge 1,
\end{eqnarray}
where $r$ is a constant exponent.
\end{lemma}
\begin{proof}
Applying Axiom 4 to resource allocation vector ${\bf 1}_{mn}$ with integers $m,n\ge 1$, we have
\begin{eqnarray}
& f({\bf 1}_{mn}) & =f(\underbrace{{\bf 1}_{m},\ldots,{\bf 1}_{m}}_{n \ {\rm segments}}) \nonumber \\
& & = f(\underbrace{m,\ldots,m}_{n \ {\rm numbers}}) \cdot  g^{-1}\left( \sum_{i=1}^n s_i\cdot g\left(f({\bf 1}_{m})\right) \right) \nonumber \\
& & =f({\bf 1}_{n})\cdot g^{-1}\left( g\left(f({\bf 1}_{m})\right) \right) \nonumber \\
& & =f({\bf 1}_{n})\cdot f({\bf 1}_{m}) \label{ap1}
\end{eqnarray}
where the second step follows from the Axiom 2 by letting $t=1/m$ and the fact that $\sum_i s_i=1$. Equation (\ref{ap1}) shows that $\log f({\bf 1}_{mn})$ is an additive number-theoretical function \cite{Erdos:46}, i.e.,
\begin{eqnarray}
\log f({\bf 1}_{mn}) = \log f({\bf 1}_{n}) + \log f({\bf 1}_{m}) \label{ap2}
\end{eqnarray}
Further, from Axiom 3, we derive
\begin{eqnarray}
\lim_{n\rightarrow \infty} \left[\log f({\bf 1}_{n+1}) - \log f({\bf 1}_{n})\right] = 0 \label{ap3}
\end{eqnarray}
Using the result in \cite{Erdos:46}, equation (\ref{ap2}) and (\ref{ap3}) implies that $ \log f({\bf 1}_{n})$ must be a logarithmic function. We have
\begin{eqnarray}
\log f({\bf 1}_{n})=r\log n,
\end{eqnarray}
where $r$ is a real constant. This is exactly (\ref{growth1}) after taking an exponential on both sides.
\end{proof}

Now, we use (\ref{growth1}) to derive an expression for the fairness measure deductively, starting from $n=2$ users. Let $x_1$ and $x_2$ be two rational numbers, such that $x_1=\frac{a_1}{b_1}$ and $x_2=\frac{a_2}{b_2}$ for some positive integers $a_1,b_1,a_2,b_2$. Using Axiom 4 and Lemma 1, we have
\begin{eqnarray}
& & (a_1b_2+a_2b_1)^r  \nonumber \\
& & \ \ \ \ = f({\bf 1}_{a_1b_2+a_2b_1})  \nonumber \\
& &  \ \ \ \ =f({\bf 1}_{a_1b_2}, {\bf 1}_{a_2b_1}) \nonumber \\
& &  \ \ \ \ =f(a_1b_2,a_2b_1)\cdot g^{-1}\left( s_1g\left(f({\bf 1}_{a_1b_2})\right) + s_2g\left(f({\bf 1}_{a_2b_1})\right) \right) \nonumber \\
& &  \ \ \ \ =f(a_1b_2,a_2b_1)\cdot g^{-1}\left( s_1g\left(a_1^rb_2^r\right) + s_2g\left(a_2^rb_1^r\right) \right) \label{ap4}
\end{eqnarray}
Applying Axiom 2 to (\ref{ap4}) with $t=b_1b_2$, we have
\begin{eqnarray}
& f(x_1,x_2) & =f\left(\frac{a_1}{b_1}, \frac{a_2}{b_2}\right) \nonumber \\
& & =f(a_1b_2,a_2b_1)  \nonumber \\
& & = \frac{(a_1b_2+a_2b_1)^r}{g^{-1}\left( s_1g\left(a_1^rb_2^r\right) + s_2g\left(a_2^rb_1^r\right) \right)} \label{ap5}
\end{eqnarray}
For a given function $g(y)$, equation (\ref{ap5}) defines fairness measure $f(x_1,x_2)$ for two users for rational vector $[x_1,x_2]$. When vector $[x_1,x_2]$ is real, by Axiom 1, fairness measure $f(x_1,x_2)$ is uniquely determined by a sequence of rational allocation vectors, whose limit is $[x_1,x_2]$. Therefore, equation (\ref{ap5}) uniquely defines fairness measure $f(x_1,x_2)$ for arbitrary real numbers $x_1,x_2$.

Suppose that we have an expression for the fairness measure $f(x_1,\ldots,x_k)$ with $k\ge 2$ users. To derive $f(x_1,\ldots,x_k,x_{k+1})$ for $k+1$ users, we use Axiom 4 to obtain the following:
\begin{eqnarray}
& & \D f(x_1,\ldots,x_k,x_{k+1}) \nonumber \\
& &  =f(\sum_{i=1}^k x_i, x_{k+1})\cdot g^{-1}\left( s_1g\left(f(x_1,\ldots,x_k)\right) + s_2g\left(f(x_{k+1})\right) \right) \nonumber \\
& &  = f(\sum_{i=1}^k x_i, x_{k+1})\cdot g^{-1}\left( s_1g\left(f(x_1,\ldots,x_k)\right) + s_2g\left(1\right) \right) \label{ap6}
\end{eqnarray}
By induction, equations (\ref{ap5}) and (\ref{ap6}) together defines fairness measure for all integer $n\ge 1$, when the mean function $g(y)$ is given. If the resulting fairness measure satisfies Axioms~1--5, it must be unique according to equations (\ref{ap5}) and (\ref{ap6}). This proves the uniqueness in Theorem 2.

To prove the existence in Theorem 1, we show that there exists a mean function $g(y)$, such that the resulting fairness index in (\ref{ap5}) and (\ref{ap6}) satisfies Axioms~1--5. We choose $g(y)=\log(y)$ and proportional weights (i.e. $\rho=1$) in (\ref{w}). From (\ref{ap5}), we derive
\begin{eqnarray}
& f(x_1,x_2) & = \frac{(a_1b_2+a_2b_1)^r}{g^{-1}\left( s_1g\left(a_1^rb_2^r\right) + s_2g\left(a_2^rb_1^r\right) \right)} \nonumber \\
& & = \frac{(a_1b_2+a_2b_1)^r}{(a_1b_2)^{rs_1}(a_2b_1)^{rs_2}} \nonumber \\
& & = \frac{(x_1+x_2)^r}{x_1^{\frac{rx_1}{x_1+x_2}}x_2^{\frac{rx_2}{x_1+x_2}}}.
\end{eqnarray}
Let $u_k=\sum_{i=1}^k x_i$ be the sum of the first $k$ elements in vector $[x_1,\ldots,x_k,x_{k+1}]$. Then, using (\ref{ap6}) inductively, we obtain
\begin{eqnarray}
& & f(x_1,\ldots,x_k,x_{k+1}) \nonumber \\
& & \ \ \ \ = f(\sum_{i=1}^k x_i, x_{k+1})\cdot g^{-1}\left( s_1g\left(f(x_1,\ldots,x_k)\right) + s_2g\left(1\right) \right) \nonumber \\
& & \ \ \ \ = f(\sum_{i=1}^k x_i, x_{k+1})\cdot f^{s_1}(x_1,\ldots,x_k) \nonumber \\
& & \ \ \ \ = \D \frac{\left(u_k +x_{k+1}\right)^r}{\left(u_k\right)^{\frac{ru_k}{u_k+x_{k+1}}}x_{k+1}^{\frac{rx_{k+1}}{u_k+x_{k+1}}}} \cdot \left[\frac{\left(u_k\right)^r}{\prod_{i=1}^kx_i^{\frac{rx_i}{u_k}}} \right]^{\frac{u_k}{u_k+x_{k+1}}} \nonumber \\
& & \ \ \ \ = \frac{\left(u_k +x_{k+1}\right)^r}{x_{k+1}^{\frac{x_{k+1}}{u_k+x_{k+1}}}\cdot\prod_{i=1}^kx_i^{\frac{rx_i}{u_k+x_{k+1}}}} \label{ap7}
\end{eqnarray}
By rearranging the terms in (\ref{ap7}), we obtain a fairness measure generated by logarithmic function $g(y)=\log(y)$ and proportional weights:
\begin{eqnarray}
f(x_1,\ldots,x_k,x_{k+1}) = \D \left(\D \sum_{i=1}^{k+1} x_i\right)^r \cdot \prod_{i=1}^{k+1}x_i^{\frac{-rx_i}{u_{k+1}}}, \label{ap8}
\end{eqnarray}
where $u_{k+1}=\sum_{i=1}^{k+1}x_i$ is the sum of all elements.

We need to prove that the fairness measure in (\ref{ap8}) satisfies Axioms~1--5. It is easy to see that Axioms~1--3 are satisfied by the fairness measure in (\ref{ap8}). To verify Axiom 4, we consider partitioning a resource allocation vector ${\bf x}$ of $n$ users into two segments: ${\bf x}^1=[x_1,\ldots,x_k]$ and ${\bf x}^1=[x_{k+1},\ldots,x_n]$ for arbitrary $0<k<n$. Let $u_{n-k}=u_n-u_k$. From (\ref{ap8}), we conclude that
\begin{eqnarray}
& & f(x_1,\ldots,x_n) \nonumber \\
& & \ = \D \left(\D \sum_{i=1}^{n} x_i\right)^r \cdot \prod_{i=1}^{n}x_i^{\frac{-rx_i}{u_{n}}}, \nonumber \\
& &  \ = \frac{(u_k+u_{n-k})^r}{u_k^{r\frac{u_k}{u_n}}u_{n-k}^{r\frac{u_n-u_k}{u_n}}}\cdot \left[\frac{u_k^r}{\prod_{i=1}^{k}x_i^{\frac{x_i}{u_k}}}\right]^{\frac{u_k}{u_n}}\cdot \left[\frac{u_{n-k}^r}{\prod_{i=k+1}^{n}x_i^{\frac{x_i}{u_{n-k}}}}\right]^{\frac{u_{n-k}}{u_n}} \nonumber \\
& & \ =f(u_k,u_{n-k})\cdot e^{\frac{u_k}{u_n}\log f({\bf x}^1)+ \frac{u_{n-k}}{u_n}\log f({\bf x}^2)} \nonumber \\
& & \ = f(\sum_{i=1}^kx_i, \sum_{i=k+1}^nx_i)\cdot  g^{-1}\left( \sum_{i=1}^2 s_i\cdot g\left(f({\bf x}^{i})\right)\right),
\end{eqnarray}
where weights $s_1=\frac{u_k}{u_n}$ and $s_2=\frac{u_{n-k}}{u_n}$ are proportional to the sum resource in each segment. This shows that the fairness measure in (\ref{ap8}) is irrelevant to partition.

To verify Axiom 5, we consider an allocation vector ${\bf x}=[\theta,1-\theta]$ and compute its fairness measure as follows
\begin{eqnarray}
f(\theta,1-\theta)=\frac{1}{\theta^{r\theta}(1-\theta)^{r(1-\theta)}}.
\end{eqnarray}
Taking a logarithm on both sides, we have
\begin{eqnarray}
\log f(\theta,1-\theta)=r\left[\theta\log\frac{1}{\theta}+(1-\theta)\log\frac{1}{1-\theta}\right]. \label{ap9}
\end{eqnarray}
Since the right hand side of (\ref{ap9}) is the entropy function, we conclude that $f(\theta,1-\theta)$ is monotonically increasing for $\theta\in[0,\frac{1}{2}]$ and monotonically decreasing for $\theta\in[\frac{1}{2},1]$. Therefore, the fairness measure in (\ref{ap8}) satisfies Axioms~1--5.

\subsection{Proof of Corollary 1}
For $n=2$ users, symmetry follows directly from equation (\ref{ap5}) in Appendix A, i.e.,
\begin{eqnarray}
f(x_1,x_2)=f(x_2,x_1), \ \forall x_1,x_2\ge 0.
\end{eqnarray}

\begin{figure*}
\begin{eqnarray}
& f(x_{i_1},\ldots,x_{i_n},x_{i_{n+1}}) &  =f\left(\sum_{j=1}^n x_{i_j}, x_{i_{n+1}}\right)\cdot  g^{-1}\left( s_1\cdot g\left(f(x_{i_1},\ldots,x_{i_n})\right) +s_2  g\left(f(x_{i_{n+1}})\right)\right) \nonumber \\
& &  =f\left(\sum_{j=1}^n x_{i_j}, x_{i_{n+1}}\right)\cdot  g^{-1}\left( s_1\cdot g\left(f(x_1,\ldots,x_{i_{n+1}-1},x_{i_{n+1}+1},\ldots,x_{n+1})\right) +s_2  g\left(f(x_{i_{n+1}})\right)\right) \nonumber \\
& &  = f(x_1,\ldots,x_{i_{n+1}-1},x_{i_{n+1}+1},\ldots,x_{n+1},x_{i_{n+1}}) \nonumber \\
& & = f\left(\sum_{j=1}^{i_{n+1}-1} x_{j}, \sum_{j=i_{n+1}}^{n+1} x_{j}\right)\cdot g^{-1}\left( s_1\cdot g\left(f(x_1,\ldots,x_{i_{n+1}-1})\right) +s_2  g\left(f(x_{i_{n+1}+1},\ldots,x_{n+1},x_{i_{n+1}})\right)\right)  \nonumber \\
& & = f\left(\sum_{j=1}^{i_{n+1}-1} x_{j}, \sum_{j=i_{n+1}}^{n+1} x_{j}\right)\cdot g^{-1}\left( s_1\cdot g\left(f(x_1,\ldots,x_{i_{n+1}-1})\right) +s_2  g\left(f(x_{i_{n+1}},x_{i_{n+1}+1},\ldots,x_{n+1})\right)\right)  \nonumber \\
& & = f(x_1,\ldots,x_{i_{n+1}-1},x_{i_{n+1}},x_{i_{n+1}+1},\ldots,x_{n+1} ) \label{eq:symmetry}
\end{eqnarray}
\end{figure*}

Assume symmetry holds for $n$ users. Let $\bx=[x_1,\ldots,x_n,x_{n+1}]$ be a resource allocation vector and $i_1,\ldots,i_n,i_{n+1}$ be an arbitrary permutation of the indices $1,\ldots,n,n+1$. When $i_{n+1}>1$, applying Axiom 4, we can use equation (\ref{eq:symmetry}) to show that
\begin{eqnarray}
f(x_{i_1},\ldots,x_{i_n},x_{i_{n+1}})=f(x_1,\ldots,x_n,x_{n+1}).
\end{eqnarray}
When $i_{n+1}=1$, using the same technique, we have
\begin{eqnarray}
& f(x_{i_1},\ldots,x_{i_n},x_{i_{n+1}}) & =f(x_{i_1},\ldots,x_{i_{n+1}},x_{i_{n}}) \nonumber \\
& & =f(x_1,\ldots,x_n,x_{n+1}).
\end{eqnarray}
Then symmetry also holds for $n+1$ users.

\subsection{Proof of Theorem 3}

Because vector ${\bf x}$ is majorized by vector ${\bf y}$, if and only if, from ${\bf x}$ we can produce ${\bf y}$ by a finite sequence of Robin Hood operations \cite{Olkin:79}, where we replace two elements $x_i$ and $x_j<x_i$ with $x_i-\epsilon$ and $x_j+\epsilon$, respectively, for some $\epsilon \in (0, x_i-x_j)$, it is necessary and sufficient to show that such an Robin Hood operation always improves a fairness measure defined by Axioms~1--5.

Toward this end, we consider partitioning a resource allocation vector ${\bf x}$ of $n$ users into two segments: ${\bf x}^1=[x_i,x_j]$ and ${\bf x}^2=[x_1,\ldots,x_{i-1},x_{i+1},\ldots,x_{j-1},x_{j+1},\ldots,x_{n}]$. Let ${\bf y}=[{\bf y}^1, {\bf x}^2]$ where ${\bf y}^1=[x_i-\epsilon, x_j+\epsilon]$ be the vector obtained from ${\bf x}^1$ by the Robin Hood operation. Using Axiom 4, we have
\begin{eqnarray}
& & \frac{f\left({\bf x}\right)}{f\left({\bf y}\right)} \nonumber \\
& &  =\frac{f({\bf x}^1,{\bf x}^2)}{f({\bf y}^1,{\bf x}^2)} \nonumber \\
& &   =\D  \frac{f(x_i+x_j, \D \sum_{ k\neq i,j} x_k)\cdot g^{-1}\left( s_1g\left(f({\bf x}^{1})\right)+s_2g\left(f({\bf x}^{2})\right)\right)}{f(x_i+x_j, \D \sum_{k\neq i,j} x_k)\cdot g^{-1}\left( s_1g\left(f({\bf y}^{1})\right)+s_2g\left(f({\bf x}^{2})\right)\right)} \nonumber \\
& &   = \frac{g^{-1}\left( s_1g\left(f(x_i,x_j)\right)+s_2g\left(f({\bf x}^{2})\right)\right)}{g^{-1}\left( s_1g\left(f(x_i-\epsilon,x_j+\epsilon)\right)+s_2g\left(f({\bf x}^{2})\right)\right)} \nonumber \\
& &  \le 1, \nonumber
\end{eqnarray}
where the last step follows form the monotonicity of $g$ and the monotonicity of fairness measure with two-users in Axiom 5, i.e.,
\begin{eqnarray}
f(x_i,x_j)\le f(x_i-\epsilon,x_j+\epsilon). \label{ap11}
\end{eqnarray}
Therefore, if ${\bf x}$ is majorized by ${\bf y}$, then we have $f({\bf x})\le f({\bf y})$. The fairness measure is Schur-concave.

\subsection{Proof of Corollary 2}
The proof for Corollary 2 is straightforward, because among the vectors with the same sum of elements, one with the equal elements is the most majorizing vector. Let $\sum_{i=1}^n x_i=n$ (which is always satisfied due to Axiom 2). The sum of the $d$ smallest elements satisfies
\begin{eqnarray}
& \sum_{i=1}^d x^{\uparrow}_i & = n\frac{\sum_{i=1}^d x^{\uparrow}_i}{\sum_{i=1}^n x^{\uparrow}_i} \nonumber \\
& & \le n\frac{d}{n} \nonumber \\
& & \le d.
\end{eqnarray}
Then, ${\bf x}\preceq {\bf 1}_n$ implies $f({\bf x})\le f({\bf 1}_n)$, for any resource allocation vector ${\bf x}$.

\subsection{Proof of Corollary 3}
Due to Schur-concavity in Theorem 3, it is sufficient to prove that collecting fixed-tax leads to a more majorizing allocation vector. From Axiom 2, we consider a vector ${\bf y}=t\left({\bf x}-c\cdot {\bf 1}_n\right)$, which achieves the same fairness as ${\bf x}-c\cdot {\bf 1}_n$, i.e.,
\begin{eqnarray}
& & f({\bf x}-c\cdot {\bf 1}_n) = f(t\left({\bf x}-c\cdot {\bf 1}_n\right))
\end{eqnarray}
where $t=\frac{\sum_i x_i}{\sum x_i-nc}$, such that
\begin{eqnarray}
\sum_{i=1}^n x_i=\sum_{i=1}^n t(x_i-c)=\sum_{i=1}^n y_i.
\end{eqnarray}
Then, for any integer $1\le d\le n$ we have
\begin{eqnarray}
& \sum_{i=1}^d y^{\uparrow}_i & = \sum_{i=1}^d t(x^{\uparrow}_i-c) \nonumber \\
& & =\frac{\sum_{i=1}^d x^{\uparrow}_i -dc}{\sum_{i=1}^n x_i-nc } \sum_{i=1}^n x_i \nonumber \\
& & \le \frac{\sum_{i=1}^d x^{\uparrow}_i -nc\frac{\sum_{i=1}^d x^{\uparrow}_i}{\sum_{i=1}^n x_i}}{\sum_{i=1}^n x_i-nc } \sum_{i=1}^n x_i  \nonumber \\
& & = \sum_{i=1}^d x^{\uparrow}_i. \nonumber
\end{eqnarray}
where the third step following from the following inequality
\begin{eqnarray}
& & \frac{\sum_{i=1}^d x^{\uparrow}_i}{\sum_{i=1}^n x_i}\le \frac{d}{n}.
\end{eqnarray}
We have ${\bf x}\succeq{\bf y}$, which implies $f({\bf x})\ge f({\bf y})=f({\bf x}-c\cdot {\bf 1})$.

\subsection{Proof of Corollary 4}
Let ${\bf x}$ be an arbitrary resource allocation vector and $t>0$ be a positive number. From Axiom 1, we have
\begin{eqnarray}
& & f({\bf x}, {\bf 0}_n) \nonumber \\
& & \ =\lim_{t\rightarrow \infty} f({\bf x}, \frac{1}{t}{\bf 1}_n) \nonumber \\
& & \ = \lim_{t\rightarrow \infty} f(\sum_{i}x_i, \frac{n}{t})\cdot g^{-1}\left( \frac{\left(\sum_ix_i\right)^{\rho}g\left(f({\bf x})\right)}{\left(\sum_ix_i\right)^{\rho}+\left(\frac{n}{t}\right)^{\rho}}\right. \nonumber \\
& & \ \ \ \ \ \ \ \ \ \ \ \ \  +\left.\frac{\left(\frac{n}{t}\right)^{\rho}g\left(f({\bf 1}_n)\right)}{\left(\sum_ix_i\right)^{\rho}+\left(\frac{n}{t}\right)^{\rho}}\right) \nonumber \\
& & \  = \lim_{t\rightarrow \infty} f(\sum_{i}x_i, \frac{n}{t})\cdot g^{-1}\left(g\left(f({\bf x}_n)\right)\right) \nonumber \\
& & \ = f({\bf x}), \nonumber
\end{eqnarray}
where the third step follows from Axiom 4.

\subsection{Proof of Theorem 4}
Without loss of generality, we assume that $f(1)=1$. First, we plug into equations (\ref{ap5}) and (\ref{ap6}) power mean $g(y)=y^\beta$ with weights generated by arbitrary $\rho$. Equation (\ref{ap5}) gives the fairness measure for two users:
\begin{eqnarray}
& f(x_1,x_2) & = \frac{(a_1b_2+a_2b_1)^r}{g^{-1}\left( s_1g\left(a_1^rb_2^r\right) + s_2g\left(a_2^rb_1^r\right) \right)} \nonumber \\
& & = \frac{(a_1b_2+a_2b_1)^r}{\left(s_1(a_1b_2)^{\beta r}+s_2(a_2b_1)^{\beta r}\right)^{\frac{1}{\beta}}} \nonumber \\
& & = \frac{(a_1b_2+a_2b_1)^r\left((a_1b_2)^{\rho}+(a_2b_1)^{\rho}\right)^{\frac{1}{\beta}}}{\left((a_1b_2)^{\rho+\beta r}+(a_2b_1)^{\rho+\beta r}\right)^{\frac{1}{\beta}}} \nonumber \\
& & = \frac{(x_1+x_2)^r\left(x_1^{\rho}+x_2^{\rho}\right)^{\frac{1}{\beta}}}{\left(x_1^{\rho+\beta r}+x_2^{\rho+\beta r}\right)^{\frac{1}{\beta}}}.
\end{eqnarray}
To derive the fairness measure for three users, we consider two different partitions of the resource allocation vector $[x_1,x_2,x_3]$ as $[x_1,x_2],[x_3]$ and $[x_1],[x_2,x_3]$. Using (\ref{ap6}), we obtain two equivalent form of the fairness measure in (\ref{th4_1}) and (\ref{th4_2}).

\begin{figure*}
\begin{eqnarray}
& f(x_1,x_2,x_3) & = f(x_1+x_2,x_3)\cdot g^{-1}\left( s_1g\left(f(x_1,x_2)\right) + s_2g\left(1\right) \right)\nonumber \\
& &  = \frac{(x_1+x_2+x_3)^r\left((x_1+x_2)^{\rho}+x_3^{\rho}\right)^{\frac{1}{\beta}}}{\left((x_1+x_2)^{\rho+\beta r}+x_3^{\rho+\beta r}\right)^{\frac{1}{\beta}}}\cdot\left[\frac{\frac{(x_1+x_2)^{\rho+\beta r}(x_1^\rho+x_2^\rho)}{x_1^{\rho+\beta r}+x_2^{\rho+\beta r}}+x_3^{\rho}}{(x_1+x_2)^{\rho}+x_3^{\rho}}\right]^{\frac{1}{\beta}} \nonumber \\
& &  = \frac{(x_1+x_2+x_3)^r}{\left((x_1+x_2)^{\rho+\beta r}+x_3^{\rho+\beta r}\right)^{\frac{1}{\beta}}}\cdot\left[\frac{(x_1+x_2)^{\rho+\beta r}(x_1^\rho+x_2^\rho)}{x_1^{\rho+\beta r}+x_2^{\rho+\beta r}}+x_3^{\rho}\right]^{\frac{1}{\beta}} \label{th4_1}
\end{eqnarray}
\begin{eqnarray}
& f(x_1,x_2,x_3) & = f(x_1,x_2+x_3)\cdot g^{-1}\left( s_1g\left(f(1)\right) + s_2g\left(x_2,x_3\right) \right)\nonumber \\
& &  = \frac{(x_1+x_2+x_3)^r\left(x_1^{\rho}+(x_2+x_3)^{\rho}\right)^{\frac{1}{\beta}}}{\left(x_1^{\rho+\beta r}+(x_2+x_3)^{\rho+\beta r}\right)^{\frac{1}{\beta}}}\cdot\left[\frac{\frac{(x_2+x_3)^{\rho+\beta r}(x_2^\rho+x_3^\rho)}{x_2^{\rho+\beta r}+x_3^{\rho+\beta r}}+x_1^{\rho}}{(x_2+x_3)^{\rho}+x_1^{\rho}}\right]^{\frac{1}{\beta}} \nonumber \\
& &  = \frac{(x_1+x_2+x_3)^r}{\left(x_1^{\rho+\beta r}+(x_2+x_3)^{\rho+\beta r}\right)^{\frac{1}{\beta}}}\cdot\left[\frac{(x_2+x_3)^{\rho+\beta r}(x_2^\rho+x_3^\rho)}{x_2^{\rho+\beta r}+x_3^{\rho+\beta r}}+x_1^{\rho}\right]^{\frac{1}{\beta}} \label{th4_2}
\end{eqnarray}
\end{figure*}

As in Axiom 4, the fairness measure is irrelevant to partition. Hence, equations (\ref{th4_1}) and (\ref{th4_2}) should be equivalent for all $x_1,x_2,x_3\ge 0$. Comparing the terms in (\ref{th4_1}) and (\ref{th4_2}), we must have $r =0$ or $\rho+\beta r=1$. When $r =0$, it is easy to see that $f(x_1,x_2,x_3)=1$ is constant. This case is trivial. We conclude that the fairness measure must have the following form
\begin{eqnarray}
f(x_1,x_2,x_3)=  \frac{\left(\sum_{i=1}^3 x_i^{1-\beta r} \right)^{\frac{1}{\beta}}}{\left(\sum_{i=1}^3 x_i \right)^{\frac{1}{\beta}-r}}, \label{th4_3}
\end{eqnarray}
where $r=\frac{1-\rho}{\beta}$ is a proper exponent. Let $u_k=\sum_{i=1}^k x_i$ be the sum of the first $k$ elements in vector $[x_1,\ldots,x_k,x_{k+1}]$. Then, using (\ref{ap6}) inductively, we obtain
\begin{eqnarray}
& & f(x_1,\ldots,x_k,x_{k+1}) \nonumber \\
& & \ = f(\sum_{i=1}^k x_i, x_{k+1})\cdot g^{-1}\left( s_1g\left(f(x_1,\ldots,x_k)\right) + s_2g\left(1\right) \right) \nonumber \\
& & \  = \D \frac{\left( u_k^{1-\beta r}+x_{k+1}^{1-\beta r} \right)^{\frac{1}{\beta}}}{\left(u_k+x_{k+1} \right)^{\frac{1}{\beta}-r}} \cdot \left[\frac{u_k^\rho \frac{\sum_{i=1}^k x_i^{1-\beta r}}{u_k^{1-\beta r}} +x_{k+1}^\rho }{u_k^\rho+x_{k+1}^\rho}\right]^{\frac{1}{\beta}} \nonumber \\
& & \ = \D \frac{\left( u_k^{1-\beta r}+x_{k+1}^{1-\beta r} \right)^{\frac{1}{\beta}}}{\left(u_k+x_{k+1} \right)^{\frac{1}{\beta}-r}} \cdot \left[\frac{\sum_{i=1}^{k} x_i^{1-\beta r} +x_{k+1}^{1-\beta r} }{u_k^{1-\beta r}+x_{k+1}^{1-\beta r}}\right]^{\frac{1}{\beta}} \nonumber \\
& & \ = \frac{\left(\sum_{i=1}^{k+1} x_i^{1-\beta r} \right)^{\frac{1}{\beta}}}{\left(\sum_{i=1}^{k+1} x_i \right)^{\frac{1}{\beta}-r}},
\label{th4_4}
\end{eqnarray}
which is exactly equation (\ref{fairness}) in Theorem 4.

We still need to prove that the fairness measure in (\ref{th4_4}) satisfies Axioms~1--5. It is easy to see that Axioms~1--3 are satisfied by the fairness measure in (\ref{th4_4}). To verify Axiom 4, we consider partitioning a resource allocation vector ${\bf x}$ of $n$ users into two segments: ${\bf x}^1=[x_1,\ldots,x_k]$ and ${\bf x}^1=[x_{k+1},\ldots,x_n]$ for arbitrary $0<k<n$. Let $u_{n-k}=u_n-u_k$ From (\ref{th4_4}), we conclude
\begin{eqnarray}
& & f(x_1,\ldots,x_n) \nonumber \\
& &  \ =  \frac{\left(\sum_{i=1}^{n} x_i^{1-\beta r} \right)^{\frac{1}{\beta}}}{\left(\sum_{i=1}^n x_i \right)^{\frac{1}{\beta}-r}} \nonumber \\
& &  \ =  \frac{\left(u_k^{1-\beta r} + u_{n-k}^{1-\beta r} \right)^{\frac{1}{\beta}}}{\left(\sum_{i=1}^n x_i \right)^{\frac{1}{\beta}-r}} \cdot\left[ \frac{ \sum_{i=1}^{n} x_i^{1-\beta r}}{u_k^{\rho} + u_{n-k}^{\rho}}  \right]^{\frac{1}{\beta}} \nonumber \\
& &  \ = f(u_k, u_{n-k})\cdot \left[ \frac{ u_k^{\rho} \frac{\sum_{i=1}^{k} x_i^{1-\beta r}}{u_k^{1-\beta r}} +u_{n-k}^{\rho} \frac{\sum_{i=k+1}^{n} x_i^{1-\beta r}}{u_{n-k}^{1-\beta r}}  }{u_k^{\rho} + u_{n-k}^{\rho}}  \right]^{\frac{1}{\beta}} \nonumber \\
& & \ = f(\sum_{i=1}^kx_i, \sum_{i=k+1}^nx_i)\cdot  g^{-1}\left( \sum_{i=1}^2 s_i\cdot g\left(f({\bf x}^{i})\right)\right),
\end{eqnarray}
where weights $s_1=\frac{u_k^\rho}{u_k^\rho+u_{n-k}^\rho}$ and $s_2=\frac{u_{n-k}^\rho}{u_k^\rho+u_{n-k}^\rho}$ are proportional to some power of the sum resource in each segment. This proves that the fairness measure in (\ref{th4_4}) is irrelevant to partition.

To verify Axiom 5, we consider an allocation vector ${\bf x}=[\theta,1-\theta]$ and compute its fairness measure as follows
\begin{eqnarray}
f(\theta,1-\theta)=\left[\theta^{1-\beta r} + (1-\theta)^{1-\beta r} \right]^{\frac{1}{\beta}}.
\end{eqnarray}
It is easy to verify that when $1-\beta r>0$, the fairness measure $f(\theta,1-\theta)$ is increasing for $\theta\in[0,\frac{1}{2}]$ and decreasing for $\theta\in[\frac{1}{2},1]$. Axiom 5 is satisfied given $1-\beta r>0$.

Putting all conditions in the proof together, we conclude that, when $\rho=1-\beta r>0$, the fairness measure given by (\ref{th4_4}) is positive and satisfies Axioms~1--5. Similarly, when $\rho=1-\beta r<0$, the fairness measure given by (\ref{fairness}) is negative. The proof for this case is the same and not repeated here.

\subsection{Proof of Corollary 5}
When $f< 0$ is negative, it is easy to show that $f({\bf x})\rightarrow -\infty$ if $x_i\rightarrow 0$. When $f>0$, suppose that $k$ users are inactive. From equation (\ref{growth1}) and Corollaries 1 and 3, we have
\begin{eqnarray}
f({\bf x}) \le f({\bf 1}_{n-k}) =n-k.
\end{eqnarray}
which gives $k\le n- f({\bf x})$. Further, since the number of active users $ n-k$ is upper bounded by $f({\bf x})$, the maximum resource is lower bounded by $\sum_i x_i/f({\bf x})$.

\subsection{Proof of Corollary 6}
Let $k({\bf x})=\sum_{i=1}^n \left(\frac{x_i}{\sum_j x_j}\right)^{1-\beta}$ be an auxiliary function, such that
\begin{eqnarray}
f({\bf x}) ={\rm sign} (1-\beta) \cdot k^{\frac{1}{\beta}}({\bf x}).
\end{eqnarray}
Since $f({\bf x})$ is differentiable, we have
\begin{eqnarray}
\frac{\partial f({\bf x})}{\partial x_i} = \frac{1}{\beta} k^{\frac{1}{\beta}-1}({\bf x}) \cdot \frac{|1-\beta |}{(\sum_j x_j)^{1-\beta}}\left[x_i^{-\beta}-\frac{\sum_j x_j^{1-\beta}}{\sum_j x_j}\right] \nonumber
\end{eqnarray}
Because $k({\bf x})>0$ is positive, $\frac{\partial f({\bf x})}{\partial x_i}$ has a single root at
\begin{eqnarray}
x_i=\bar{x}=\left(\frac{\sum_j x_j}{\sum_j x_j^{1-\beta}}\right)^{\frac{1}{\beta}}.
\end{eqnarray}
It is straightforward to show that for any $\beta \neq 1$, we have
\begin{eqnarray}
\frac{\partial f({\bf x})}{\partial x_i} >0, \ {\rm if} \ x_i>\bar{x} \ \  {\rm and} \ \
\frac{\partial f({\bf x})}{\partial x_i} <0, \ {\rm if} \ x_i<\bar{x} \nonumber
\end{eqnarray}
Therefore, when $x_j$ $\forall j\neq i$ are fixed, $f({\bf x})$ is maximized by $x_i=\bar{x}$.

\subsection{Proof of Corollary 7}
To derive an lower bound on $f({\bf x})$ under the box constraints $x_{min}\le x_i\le x_{max}$ $\forall i$, we first argue that $f({\bf x})$ is minimized only if users are assigned resource $x_{min}$ or $x_{max}$. Using the box constraints and Corollary 6, we have
\begin{eqnarray}
& \bar{x} & =\left(\frac{\sum_j x_j}{\sum_j x_j^{1-\beta}}\right)^{\frac{1}{\beta}} \nonumber \\
& & = \left(\sum_i \frac{x_i}{\sum_j x_j}\cdot x_i^{-\beta}\right)^{-\frac{1}{\beta}} \nonumber \\
& & \ge \left(\sum_i \frac{x_i}{\sum_j x_j}\cdot x_{min}^{-\beta}\right)^{-\frac{1}{\beta}}  \nonumber \\
& & = x_{min}.
\end{eqnarray}
Similarly, we can show
\begin{eqnarray}
\bar{x} \le  x_{max}.
\end{eqnarray}
According to Axiom 4, $f({\bf x})$ is increasing on $x_i\in[x_{min},\bar{x}]$ and decreasing on $x_i\in[\bar{x},x_{max}]$. Hence, $f({\bf x})$ is minimized only if all $x_i$ take the boundary values in the box constraints, i.e.,
\begin{eqnarray}
x_i=x_{min} \ {\rm or} \ x_i=x_{max}.
\end{eqnarray}

Let $\Gamma=\frac{x_{max}}{x_{min}}$ and $\mu$ be fraction of users who receive $x_{max}$. By relaxing the constraint $\mu\in\left\{\frac{i}{n},\forall i\right\}$ to $\mu\in[0,1]$, we derive an lower bound on $f({\bf x})$ as follows
\begin{eqnarray}
& & \min_{x_i\in [x_{min},x_{max}],\forall i} f({\bf x}) \nonumber \\
& & \ \ = \min_{\mu\in\left\{\frac{i}{n},\forall i\right\}} {\rm sign}(1-\beta)\cdot n\left[\frac{ \mu\Gamma^{1-\beta}+(1-\mu)}{\left(\mu\Gamma+1-\mu\right)^{1-\beta}}\right]^{\frac{1}{\beta}} \nonumber \\
& & \ \ \ge \min_{\mu\in[0,1]} {\rm sign}(1-\beta)\cdot n\left[\frac{ \mu\Gamma^{1-\beta}+(1-\mu)}{\left(\mu\Gamma+1-\mu\right)^{1-\beta}}\right]^{\frac{1}{\beta}}. \label{ap12}
\end{eqnarray}
To find the minimizer in the last optimization problem above, we first recognize that at the two boundary points $\mu=0$ and $\mu=1$ (i.e. all users receive the same amount of resource), $f({\bf x})=n$ achieves its maximum value. Therefore, the minimum value is achieved by some $\mu\in(0,1)$. If $\mu^*$ is the minimizer of (\ref{ap12}), it is necessary that the first order derivative of the right hand side of (\ref{ap12}) is zero, i.e.,
\begin{eqnarray}
\frac{\partial \left[\frac{ \mu\Gamma^{1-\beta}+(1-\mu)}{\left(\mu\Gamma+1-\mu\right)^{1-\beta}}\right]}{\partial \mu}=0.
\end{eqnarray}
Soling the above equation, we obtain
\begin{eqnarray}
(\Gamma-1)(1-\beta)\left[\left(\Gamma^{1-\beta}-1\right)\mu+1\right]=\left(\Gamma^{1-\beta}-1\right)\left[(\Gamma -1)\mu+1\right]. \nonumber
\end{eqnarray}
Because this equation is a linear in $\mu$, its root $\mu^*$ is the unique minimizer of (\ref{ap12}):
\begin{eqnarray}
\mu^*=\frac{\Gamma-\Gamma^{1-\beta}-\beta(\Gamma-1)}{\beta(\Gamma-1)(\Gamma^{1-\beta}-1)}.
\end{eqnarray}
The lower bound in Corollary 7 follows by plugging $\mu^*$ into (\ref{ap12}).

\subsection{Proof of Theorem 5}
We first prove the monotonicity of $f_\beta ({\bf x})$ for $\beta\in (-\infty,0)$. Consider two different values $0>\beta_1\ge\beta_2$. We define the a function $\phi(y)=y^{\frac{\beta_2}{\beta_1}}$ for $y\in \mathbb{R}_+$. Since $\beta_2/\beta_1 \ge 1$, the function $\phi(y)$ is convex in $y$. Therefore, we have
\begin{eqnarray}
& f_{\beta_2} ({\bf x}) & = \left[\sum_{i=1}^n \left(\frac{x_i}{\sum_j x_j}\right)^{1-\beta_2}\right]^{\frac{1}{\beta_2}}  \nonumber \\
& & = \left[\sum_{i=1}^n  \frac{x_i}{\sum_j x_j} \cdot \phi\left(\left(\frac{x_i}{\sum_j x_j}\right)^{-\beta_1}\right)\right]^{\frac{1}{\beta_2}} \nonumber \\
& & \le \left[\phi\left(\sum_{i=1^n}\frac{x_i}{\sum_j x_j} \left(\frac{x_i}{\sum_j x_j}\right)^{-\beta_1} \right)\right]^{\frac{1}{\beta_2}} \nonumber \\
& & = \left[\phi\left(\sum_{i=1}^n  \left(\frac{x_i}{\sum_j x_j}\right)^{1-\beta_1} \right)\right]^{\frac{1}{\beta_2}} \nonumber \\
& & = \left[\sum_{i=1}^n  \left(\frac{x_i}{\sum_j x_j}\right)^{1-\beta_1} \right]^{\frac{1}{\beta_1}} \nonumber \\
& & = f_{\beta_2} ({\bf x}),
\end{eqnarray}
where the third step follows from Jensen's inequality and $\beta_2<0$. This shows that $f_\beta ({\bf x})$ is increasing on $(-\infty,0)$.

For $\beta\in (0,1)$, we consider $1>\beta_1\ge\beta_2>0$. The function $\phi(y)=y^{\frac{\beta_2}{\beta_1}}$ becomes concave. We have
\begin{eqnarray}
& f_{\beta_2} ({\bf x}) & = \left[\sum_{i=1}^n \left(\frac{x_i}{\sum_j x_j}\right)^{1-\beta_2}\right]^{\frac{1}{\beta_2}}  \nonumber \\
& & = \left[\sum_{i=1}^n  \frac{x_i}{\sum_j x_j} \cdot \phi\left(\left(\frac{x_i}{\sum_j x_j}\right)^{-\beta_1}\right)\right]^{\frac{1}{\beta_2}} \nonumber \\
& & \le \left[\phi\left(\sum_{i=1^n}\frac{x_i}{\sum_j x_j} \left(\frac{x_i}{\sum_j x_j}\right)^{-\beta_1} \right)\right]^{\frac{1}{\beta_2}} \nonumber \\
& & = f_{\beta_2} ({\bf x}).
\end{eqnarray}
where the third step follows from Jensen's inequality and $\beta_2>0$. Therefore,  $f_\beta ({\bf x})$ is increasing on $(0,1)$.

For $\beta\in (1,\infty)$, we consider $\beta_1\ge\beta_2>1$. The function $\phi(y)=y^{\frac{\beta_2}{\beta_1}}$ is concave. We have
\begin{eqnarray}
& f_{\beta_2} ({\bf x}) & -\left[\sum_{i=1}^n \left(\frac{x_i}{\sum_j x_j}\right)^{1-\beta_2}\right]^{\frac{1}{\beta_2}}  \nonumber \\
& & = -\left[\sum_{i=1}^n  \frac{x_i}{\sum_j x_j} \cdot \phi\left(\left(\frac{x_i}{\sum_j x_j}\right)^{-\beta_1}\right)\right]^{\frac{1}{\beta_2}} \nonumber \\
& & \ge -\left[\phi\left(\sum_{i=1^n}\frac{x_i}{\sum_j x_j} \left(\frac{x_i}{\sum_j x_j}\right)^{-\beta_1} \right)\right]^{\frac{1}{\beta_2}} \nonumber \\
& & = f_{\beta_2} ({\bf x}).
\end{eqnarray}
where the third step follows from Jensen's inequality and $\beta_2>0$. Therefore,  $f_\beta ({\bf x})$ is decreasing on $(1,\infty)$. This completes the proof of Theorem 5.

% \section{Proof of Theorem 6}
%  When $\alpha=\beta>1$, using Theorem 3 and 5, we obtain
%  \begin{eqnarray}
%  & \frac{\partial |f_{\beta}({\bf x})|^\beta}{\partial \beta} & = \frac{\partial \left[-f_{\beta}({\bf x})\right]^\beta}{\partial \beta} \nonumber \\
%  & & = -\beta\left[-f_{\beta}({\bf x})\right]^{\beta-1}\frac{\partial f_{\beta}({\bf x})}{\partial \beta} \nonumber \\
%  & & \ge 0
%  \end{eqnarray}
%  where the second step holds because $f_{\beta}({\bf x})$ is negative and monotonically decreasing on $(1,\infty)$.
%  For the efficiency component, we have
%  \begin{eqnarray}
%  & \frac{\partial |U_{\beta}({\bf x})|^\beta}{\partial \beta} & = \frac{\partial {\left(\sum_i x_i\right)^{1-\beta}}/{(\beta-1)}}{\partial \beta} \nonumber \\
%  & & = \frac{-\left(\sum_i x_i\right)^{1-\beta}\log\left(\sum_i x_i\right)(\beta-1)-\left(\sum_i x_i\right)^{1-\beta}}{(\beta-1)^2} \nonumber \\
%  & & \le 0
%  \end{eqnarray}
%  where the second step holds because $\sum_i x_i>1$ according to the assumption in Theorem6. Therefore, the tradeoff between fairness and efficiency is proven.

\subsection{Proof of Theorem 6}

We first assume $\beta>1$ (which implies $f_\beta(\cdot)<0$) and show that the condition $\lambda\le \left|\frac{\beta}{1-\beta}\right|$ is necessary and sufficient for preserving Pareto optimality. The case where $\beta<1$ can be shown using a completely analogous proof.

To show that the condition $\lambda\le \left|\frac{\beta}{1-\beta}\right|$ is sufficient, we consider an allocation ${\bf x}$ and a vector $\boldsymbol{\gamma}$ such that $\gamma_i \geq 0$ for all $i$ and $\sum_i\gamma_i = \sum_i x_i$. Clearly, $\mathbf{x}' = {\bf x} + \delta\boldsymbol{\gamma}$ Pareto dominates $\mathbf{x}$ for $\delta>0$. We now consider the difference between the function (\ref{eq:tradeoff}) evaluated for these two allocations. First assume $\beta>1$, which implies $f_\beta(\cdot)<0$, and
\begin{align}
	\Phi_{\lambda}(\mathbf{x}')& - \Phi_{\lambda}(\mathbf{x}) \nonumber \\
		={}& \lambda \left(\ell\left(f_\beta\left(\mathbf{x}'\right)\right)-\ell\left(f_\beta\left(\mathbf{x}\right)\right)\right) + \ell\left(\sum_i x'_i\right)-\ell\left(\sum_i x_i\right) \nonumber  \\
		={}& -\lambda \left(\log\left|f_\beta\left(\mathbf{x}'\right)\right|-\log\left|f_\beta\left(\mathbf{x}\right)\right|\right) + \log\left((1+\delta)\sum_i x_i\right) \nonumber \\
{} & \ \ \ \ \ -\log\left(\sum_i x_i\right) \nonumber  \\
		={}& -\lambda \left(\log\left|f_\beta\left(\mathbf{x}'\right)\right|-\log\left|f_\beta\left(\mathbf{x}\right)\right|\right) + \log\left(1+\delta\right).\label{rst0}
\end{align}
If $\mathbf{x}'$ is also more fair than $\mathbf{x}$, then showing
\begin{equation}
	-\lambda \left(\log\left|f_\beta\left(\mathbf{x}'\right)\right|-\log\left|f_\beta\left(\mathbf{x}'\right)\right|\right) > 0
\end{equation}
is trivial, and the difference between the objective evaluated at the two allocations is strictly positive. Therefore, we consider the case where $\mathbf{x}'$ is less fair.

Continuing from (\ref{rst0}) and applying the definition in (\ref{r0}) yields
\begin{align}
	\Phi_{\lambda}(\mathbf{x}')& - \Phi_{\lambda}(\mathbf{x}) \nonumber \\
	={}& -\lambda \log\left(\left[\sum_{i=1}^n \left(\frac{x'_i}{\sum_j x'_j}\right)^{1-\beta} \right]^{\frac{1}{\beta}}\right)+ \log\left(1+\delta\right) \nonumber \\
{}& \ \ \ \ \ +\lambda \log\left(\left[\sum_{i=1}^n \left(\frac{x_i}{\sum_j x_j}\right)^{1-\beta} \right]^{\frac{1}{\beta}}\right)  \nonumber \\
	={}& -\frac{\lambda}{\beta} \log\left(\frac{\sum_{i=1}^n \left(x'_i\right)^{1-\beta} }{\sum_{i=1}^n \left(x_i\right)^{1-\beta} }\right) + \left(1 - \lambda\frac{\beta-1}{\beta}\right)\log\left(1+\delta\right). \label{rst01}
\end{align}

Because $x'_i\geq x_i$ for all $i$, we know that for $\beta>1$, $(x'_i)^{1-\beta}\leq (x_i)^{1-\beta}$, which implies
\begin{equation}
	-\frac{\lambda}{\beta}\log\left(\frac{\sum_{i=1}^n \left(x'_i\right)^{1-\beta}}{\sum_{i=1}^n \left(x_i\right)^{1-\beta}}\right) > 0.
\end{equation}

Consequently, for the entire difference to be positive, it is sufficient that
\begin{equation}
	1 - \lambda\frac{\beta-1}{\beta}\geq 0,
\end{equation}
or, equivalently,
\begin{equation}
	\lambda\leq \frac{\beta}{\beta-1}.
\end{equation}

Next, we prove that the condition $\lambda\le \left|\frac{\beta}{1-\beta}\right|$ is necessary. Suppose $\beta>1$ and $\lambda > \left|\frac{\beta}{1-\beta}\right|$. We show that there exists two vectors ${\bf x}$ and ${\bf x'}$, such that $\Phi_{\lambda}(\mathbf{x}') - \Phi_{\lambda}(\mathbf{x})<0$, while $\mathbf{x}'$ Pareto dominates $\mathbf{x}$.

Consider an allocation ${\bf x}$ of length $n+1$, such that $x_i=1$ for $i=1,\ldots,n$ and $x_{n+1}=n$. Clearly, $\mathbf{x}$ is Pareto dominated by another vector ${\bf x}'$, where $x_i'=x_i$ for $i=1,\ldots,n$ and $x_{n+1}'=x_i+\delta\left(\sum_{i} x_i\right)$, for some positive $\delta>0$. From the last step of (\ref{rst01}), we have
\begin{align}
	\Phi_{\lambda}(\mathbf{x}')& - \Phi_{\lambda}(\mathbf{x}) \nonumber \\
	={}& -\frac{\lambda}{\beta} \log\left(\frac{\sum_{i=1}^{n+1} \left(x'_i\right)^{1-\beta} }
		{\sum_{i=1}^{n+1} \left(x_i\right)^{1-\beta} }\right)
		+ \left(1 - \lambda\frac{\beta-1}{\beta}\right)\log\left(1+\delta\right) \nonumber \\
	={}& -\frac{\lambda}{\beta} \log\left(\frac{n+ \left(n+2n\delta\right)^{1-\beta} }{n+n^{1-\beta} }\right) + \left(1 - \lambda\frac{\beta-1}{\beta}\right)\log\left(1+\delta\right) \nonumber \\
	\le {}& -\frac{\lambda}{\beta} \log\left(\frac{n }{n+n^{1-\beta} }\right) + \left(1 - \lambda\frac{\beta-1}{\beta}\right)\log\left(1+\delta\right) \nonumber \\
	= {}& -\frac{\lambda}{\beta} \log\left(1+n^{-\beta} \right) + \left(1 - \lambda\frac{\beta-1}{\beta}\right)\log\left(1+\delta\right) \nonumber
\end{align}
It is straight forward to verify that $\Phi_{\lambda}(\mathbf{x}') - \Phi_{\lambda}(\mathbf{x})<0$, if we set
\begin{equation}
	\delta=\frac{1}{2}\left[\left(1+n^{-\beta} \right)^{\frac{\lambda}{\lambda(\beta-1)-\beta}}\right]>0.
\end{equation}

As a result, the condition (\ref{eq:paretocond}) of the theorem is sufficient and necessary for ensuring Pareto optimality of the solution.

\subsection{Proof of Theorem 7}

From the definition of $\alpha$-fair utility, we compute the numerator and denominator:
\begin{align}
	 \left\langle\bigtriangledown_\mathbf{x}U_{\alpha=\beta}(\mathbf{x}),\frac{\boldsymbol{\eta}}{\|\boldsymbol{\eta}\|}\right\rangle
		={}&\sum_{i} x_i^{-\beta}\frac{\frac{1}{N}-\frac{x_i}{\sum_i x_i}}{\sqrt{\frac{\|\mathbf{x}\|^2}{(\sum_i x_i)^2}-\frac{1}{N}}} \nonumber \\
		={}&\frac{1}{\sqrt{\frac{\|\mathbf{x}\|^2N}{(\sum_i x_i)^2}-1}}\cdot\frac{1}{\sqrt{N}} \nonumber \\
		& \ \ \ 	\cdot\sum_{i} x_i^{-\beta}\left(1-\frac{x_i}{\sum_j x_j}N\right),
\end{align}
and
\begin{align}
	\left\langle\displaystyle\bigtriangledown_\mathbf{x} U_{\alpha=\beta}(\mathbf{x}),\frac{\mathbf{1}}{\|\mathbf{1}\|}\right\rangle
		={}& \sum_{i} x_i^{-\beta}\frac{1}{\sqrt{N}}\\
		={}& \frac{1}{\sqrt{N}}\sum_{i} x_i^{-\beta}.
\end{align}
Notice that both values are positive. The ratio between these then is
\begin{align}
	 \frac{\left\langle\bigtriangledown_\mathbf{x}U_{\alpha=\beta}(\mathbf{x}),\frac{\boldsymbol{\eta}}{\|\boldsymbol{\eta}\|}\right\rangle}
		{\left\langle\displaystyle\bigtriangledown_\mathbf{x} U_{\alpha=\beta}(\mathbf{x}),\frac{\mathbf{1}}{\|\mathbf{1}\|}\right\rangle}
		={}&\frac{1}{\sqrt{\frac{\|\mathbf{x}\|^2N}{(\sum_i x_i)^2}-1}}\left(1 - \frac{\sum_i\frac{x_i}{\sum_jx_j}x_i^{-\beta}}{\sum_i\frac{1}{N}x_i^{-\beta}}\right).
\end{align}
It is easily shown that the factor out fron is strictly positive. The only component that varies with $\beta$ is the ratio between two weighted averages of the same vector with different weights:
\begin{equation}
	\frac{\sum_i\frac{x_i}{\sum_jx_j}x_i^{-\beta}}{\sum_i\frac{1}{N}x_i^{-\beta}}.\label{eq:mean_ratio}
\end{equation}
That average in the numerator places more weight ($\frac{x_i}{\sum_jx_j} > \frac{1}{N}$) on elements that decrease more rapidly (or increase more slowly for the case $x_i<1$) with $\beta$, implies that the overall numerator decreases more rapidly (or increases more slowly) than the denominator. Therefore, (\ref{eq:mean_ratio}) is monotonically non-increasing, and Theorem~\ref{thm:fairer} is true.

\subsection{Proof of Theorem 8}
To prove Theorem 8, we need to show that if $F({\bf x})$ satisfies Axioms~1$^\prime$-4$^\prime$, its normalization
\begin{equation}
f({\bf x})=F({\bf x})\cdot\left(\sum_i x_i\right)^{-\frac{1}{\lambda}} \label{normalization}
\end{equation}
is a fairness measure satisfying Axioms~1-5.

The continuity of $f({\bf x})$ follows directly from that of $F({\bf x})$ in Axioms~1$^\prime$. Let $z>0$ be a positive real number and ${\bf y}$ be a vector of arbitrary length. To prove homogeneity, we make use of Axioms~3$^\prime$:
\begin{eqnarray}
& & F(z\cdot[{\bf y},{\bf y}]) \nonumber \\
& & \ \ \ \ = F\left(z,z\right)\cdot  g^{-1}\left( s_1\cdot g\left(F({\bf y})\right) + s_2\cdot g\left(F({\bf y})\right)\right) \nonumber \\
& & \ \ \ \ = F\left(z,z\right)\cdot  F({\bf y}) \nonumber \\
& & \ \ \ \ = F(1,1) \cdot g^{-1}\left( s_1\cdot g\left(F(z)\right) + s_2\cdot g\left(F(z)\right)\right) \cdot F({\bf y}) \nonumber \\
& & \ \ \ \ = F(1,1) \cdot F({\bf y})  \cdot F(z)  \nonumber
\end{eqnarray}
and similarly,
\begin{eqnarray}
& & F(z{\bf y},z{\bf y}) \nonumber \\
& & \ \ \ \ = F\left(1,1\right)\cdot   g^{-1}\left( s_1\cdot g\left(F(z{\bf y})\right) + s_2\cdot g\left(F(z{\bf y})\right)\right)  \nonumber \\
& & \ \ \ \ = F\left(1,1\right)\cdot F(z{\bf y})
\end{eqnarray}
Comparing the above two equations, we have
\begin{eqnarray}
& & F(z{\bf y})  = F(z) \cdot F({\bf y})   . \label{produ}
\end{eqnarray}
When ${\bf y}$ is a scalar, using the result in \cite{Erdos:46}, equation (\ref{produ}) implies that $\log F(z)=\frac{1}{\lambda}\log(z)$ must be a logarithmic function with an exponent $\frac{1}{\lambda}$. We have
\begin{eqnarray}
& & F(z{\bf y})  = z^{\frac{1}{\lambda}} F({\bf y}),
\end{eqnarray}
which is a homogenous function of order $\frac{1}{\lambda}$. Therefore, its normalization $f({\bf x})$ in (\ref{normalization}) is a homogenous function of order zero and satisfies Axiom~2.

Using the homogeneity property and Axioms~2$^\prime$, we obtain
\begin{eqnarray}
& \lim_{n\rightarrow \infty} \frac{f({\bf 1}_{n+1})}{f({\bf 1}_{n})} & = \lim_{n\rightarrow \infty} \frac{F({\bf 1}_{n+1})}{F({\bf 1}_{n})}\cdot \left(1+\frac{1}{n}\right)^{-\frac{1}{\lambda}} \nonumber \\
& & = \lim_{n\rightarrow \infty} \frac{F({\bf 1}_{n+1})}{F({\bf 1}_{n})} \nonumber \\
& & =1.
\end{eqnarray}
This is exactly Axiom~3. From Axiom 4$^\prime$m, Axiom 5 is straight since
\begin{eqnarray}
f(\theta,1-\theta) = F(\theta,1-\theta)\cdot(\theta+1-\theta)^{-\frac{1}{\lambda}}=F(\theta,1-\theta). \nonumber
\end{eqnarray}
Therefore, monotonicity holds for $f(\theta,1-\theta)$ for $\theta\in[0,\frac{1}{2}]$ and $\theta\in[0,\frac{1}{2}]$, respectively. To prove Axiom~4, we choose $x_1=w({\bf y}^1)$ and $x_2=w({\bf y}^2)$ in Axiom 3$^\prime$, which results in \begin{eqnarray}
& & f({\bf y^1},{\bf y^2}) \nonumber \\
& & \ \ \ =F({\bf y^1},{\bf y^2})\cdot\left(w({\bf y}^1)+w({\bf y}^2)\right)^{-\frac{1}{\lambda}} \nonumber \\
& & \ \ \  =F\left(x_1,x_2\right)\cdot  g^{-1}\left( \sum_{i=1}^2 s_i\cdot g\left(F({\bf y}^{i}/x_i)\right) \right) \cdot\left(x_1+x_2\right)^{-\frac{1}{\lambda}} \nonumber \\
& & \ \ \ =F\left(x_1,x_2\right)\cdot\left(x_1+x_2\right)^{-\frac{1}{\lambda}} \cdot  g^{-1}\left( \sum_{i=1}^2 s_i\cdot g\left(f({\bf y}^{i}/x_i)\right) \right) \nonumber \\
& & \ \ \ =f\left(x_1,x_2\right) \cdot  g^{-1}\left( \sum_{i=1}^2 s_i\cdot g\left(f({\bf y}^{i})\right) \right) \nonumber \\
& & \ \ \ =f\left(w({\bf y}^1,w({\bf y}^2\right) \cdot  g^{-1}\left( \sum_{i=1}^2 s_i\cdot g\left(f({\bf y}^{i})\right) \right)
\end{eqnarray}
where the second last step uses the fact that $f$ is a homogenous function of order zero. This proves Axiom~4.

If $F({\bf x})$ satisfies Axioms~1$^\prime$-4$^\prime$, we have shown that its normalization $f(x)$ is a fairness measure satisfying Axioms~1-5. Therefore, $F(x)$ is homogenous function of order $\frac{1}{\lambda}$ and is given by
\begin{equation}
F({\bf x})=f({\bf x})\cdot\left(\sum_i x_i\right)^{\frac{1}{\lambda}}.
\end{equation}
Existence and unique of $F({\bf x})$ is straightforward from that of $f({\bf x})$ in Theorems 1 and 2.

\end{document}